\pdfoutput=1
\documentclass[11pt]{article}
\usepackage{amssymb}
\usepackage{graphicx}
\usepackage{amsmath}
\usepackage{parskip}
\newcommand{\ol}{\overline}
\newcommand{\es}{\emptyset}
\newcommand{\ra}{\rightarrow}
\newcommand{\Ra}{\Rightarrow}

\newcommand{\s}{\subseteq}

\title{Compression with wildcards: All models of a Boolean 2-CNF}

\author{Marcel Wild}
\date{}
\begin{document}
\maketitle

\begin{quote}
{\bf Abstract. }{\sl Let $W$ be a finite set which simultaneously serves as the universe of any poset $(W,\le)$ and as the vertex set of any graph $G$. Our algorithm, abbreviated\\ A-I-I, enumerates  all $G$-independent  ideals of $(W,\le)$.
Since every satisfiable Boolean 2-CNF can be Horn-renamed,  A-I-I becomes the core of  a polynomial total time algorithm that enumerates the modelset of  any Boolean 2-CNF. As a perk, the modelset is delivered in a compressed format (that uses don't-care symbols).}
\end{quote}


\section{Introduction}

We recommend [CH] for an introduction to Boolean functions and for reading up all undefined terms in the present article. A {\it Boolean 2-CNF} is a Boolean function $F$ with vector of variables $x=(x_1,x_2,...,x_w)$ that is given in conjunctive normal form and that has merely clauses of length at most  two. We recommend [S]  (a chapter within [CH]) for a survey of the many applications of 2-CNFs, both outside and within mathematics. Additionally the reader may wish to inspect [Ma] which is dedicated to the immensely applicable stable matchings and their tight connection to 2-CNFs. So much for applications; the author will henceforth concentrate on the purely mathematical and algorithmic aspects of 2-CNFs.

Any $y\in\{0,1\}^w$ with $F(y)=1$ is a {\it model} of $F$. Let $Mod(F)\s\{0,1\}^w$ be the set of all $F$-models. For instance, letting $w=4$ consider

$$F_1(x)=(\ol{x_1}\vee x_3)\wedge (\ol{x_1}\vee x_4)\wedge (\ol{x_2}\vee x_1)\wedge (\ol{x_2}\vee \ol{x_4})\wedge (x_2\vee x_4)\wedge (\ol{x_3}\vee x_4)\wedge  \ol{x_2}.$$

To warm up, notice that the 1-clause $\ol{x_2}$ forces $y_2=0$ for each model $y$ of $F_1$. But this imlies that $F_1$ boils down\footnote{While the domains of $F_1$ and $F_2$ are $\{0,1\}^4$ and $\{0,1\}^3$ respectively (and so $F_1\neq F_2$), it holds that $Mod(F_1)=\{(y_1,0,y_3,y_4):\ (y_1,y_3,y_4)\in Mod(F_2)\}$.} to
$F_2(\vec{x})=(\ol{x_1}\vee x_3)\wedge (\ol{x_1}\vee x_4)\wedge x_4\wedge (\ol{x_3}\vee x_4).$
This further implies $y_4=1$ for each model $y$ of $F_1$, and so $F_2$ further boils down to $F_3(\vec{x})=(\ol{x_1}\vee x_3)$.

 From this it is clear that each satisfiable\footnote{The {\it Labelling Algorithm} of [CH,p.231], which works along lines similar to the above toy example, decides in linear time whether or not a given 2-CNF is satisfiable. It will be crucial later that deciding 2-SAT is much easier than deciding SAT (=satisfiability) for arbitrary CNFs.} 2-CNF is easily reduced to a 2-CNF $F$ without 1-clauses. All 2-CNFs are henceforth silently assumed to be of this kind. Consequently each clause is either
  {\it positive} ($x_i\vee x_j$), or {\it negative} ($\ol{x_i}\vee \ol{x_j}$),
or {\it mixed}  ($x_i\vee \ol{x_j}$).

\vspace{5mm}
 Our quest is to enumerate {\it all} models of a 2-CNF $F$. The pioneering and so far only work in this regard is the 1994 article [F] of Feder. 
 After some preprocessing time proportional to the number of clauses, Feder's algorithm outputs the models in $O(w)$ time per model\footnote{More specifically in time $O(d)$ where $d$ is some parameter whose most natural upper bound is $w$. We note that Feder speaks of 'solutions' rather than 'models'.}, and whence qualifies as a polynomial-delay (even linear-delay) enumeration algorithm. In contrast our enumeration method, called {\tt All-Independent-Ideals} (A-I-I), cannot boast polynomial-delay, just polynomial total time $O(Rw^3)$ (where $R$ is explained in a moment). 
 
 Nevertheless, we feel A-I-I compares favorably to Feder's algorithm in three aspects. First,
 it enumerates $Mod(F)$ in a {\it compressed} format that uses don't-care symbols '2' which can be freely substituted by 1-bits or 0-bits. Thus $Mod(F)$ is output as a union of $R$  mutually disjoint {\it 012-rows} of length $w$, such as $(2,1,1,2,0,1,2,1,0,2)$ (for $w=10$) which comprises $2^4$ models. Of course, if $Mod(F)$  runs into the billions, and compression is high, then even constant-delay algorithms will trail A-I-I. 
 Second, the efficiency of A-I-I is {\it evidenced} by comparison to Mathematica's state of the art command {\tt BooleanConvert}. Third, A-I-I is easy to {\it parallelize},
 and so in principle can be sped-up to any desired extent.

 \vspace{3mm}
 Actually {\tt All-Independent-Ideals}  (discussed in Sec.2) only applies to some (a priori) rather specific  2-CNFs, i.e. to acyclic Horn 2-CNFs $H(x)$. That $H(x)$ is {\it Horn} means that all its clauses are either negative or mixed (but not positive). The definition of "acyclic" is postponed.

 How can arbitrary 2-CNFs $F(x)$ be reduced to acyclic Horn  2-CNFs? For starters, recall that the satisfiability of $F(x)$ is easy to determine. If $F$ is insatisfiable then $Mod(F)=\es$. Otherwise any model of $F$ yields (Sec.5) some "magic" set of variables, say $\{x_2,x_7,x_{19}\}$, with the following property. Replacing each occurence of $x_2$ in $F$ by $\ol{x}_2$, and conversely each $\ol{x}_2$ by $x_2$, and likewise for $x_7,x_{19}$, yields a new formula $H'$ which is a Horn 2-CNF.
 
 When the Horn 2-CNF $H'$  features directed cycles in some associated digraph, the latter are "factored out" and the result (Sec.4) is an {\it acyclic} Horn 2-CNF $H$.
 The transition from $F$ to $H$ is fast, and $Mod(H)$ (calculated by A-I-I) immediately yields $Mod(F)$.  
 
 We implemented A-I-I in high-level Mathematica code and compared it (Sec. 3) with the  Mathematica command {\tt BooleanConvert}. The latter is a general purpose routine that enumerates (also in compressed format) the models of any Boolean function. Depending on the type of input, one or the other of the two prevails.

 \vspace{3mm}
 In Section 6 we ponder the introduction of wildcards beyond the don't-care symbol.
 This usually increases compression when $F$ is homogeneous in the sense of having exclusively mixed, or negative, or positive clauses. In all other cases the picture is not so clear-cut and readers are encouraged to code some of the proposed ideas.

\vspace{3mm}
Section 7 accompanies all these ideas with a brief survey of general All-SAT strategies (which are thus not focused on 2-CNFs).
We will use the shorthand "iff" for "if and only if".

\section{Compressing all independent order ideals}

The core Subsections 2.3.1 and 2.4 are soly about graphs and posets on a common universe, and hence could be read without knowing anything about Boolean functions. However, in view of things to come it seemed a good idea to expose the reader to Boolean functions beforehand.

A Boolean 2-CNF is {\it mixed} if  all its clauses are mixed (as defined in Sec.1), and it is {\it negative} if all its clauses are negative. Roughly speaking the two kinds correspond  to posets (2.1) and graphs (2.2) respectively. In 2.3 (toy example) and 2.4 (detailed algorithm) the two structures {\it share a common universe}. We strive to enumerate all (poset) ideals that happen to be independent in the graph.

\vspace{5mm}
{\bf 2.1} As to mixed 2-CNFs, consider $H_{mix,1}:\{0,1\}^8\ra\{0,1\}$ defined by

$$(1)\quad H_{mix,1}(x):=(\ol{x_8}\vee x_1)\wedge(\ol{x_6}\vee x_2)\wedge(\ol{x_7}\vee x_2)\wedge(\ol{x_7}\vee x_3)
\wedge(\ol{x_5}\vee x_3)$$

\noindent
Since  $\ol{x_i}\vee x_j$ is logically equivalent to  $x_i\ra x_j$, the models of $H_{mix,1}$ match the  ideals of the poset $(W_1,\le)$ in Figure 1. For instance $x_7\to x_2$ by definition\footnote{This definition (as opposed to ($x_7<x_2$) suits us better in view of [W2].} implies $x_2<x_7$.
Do not confuse the partial order relation $\le$ with the ordinary total ordering of natural numbers, denoted by $\le'$. In fact the two are interwoven in that
the labeling $1,2,..,8$ of the elements of $W_1$ is a so-called\footnote{Older texts speak of 'topological orderings'.} {\it linear extension} of $W_1$, i.e.

$$(2)\quad  (\forall i,j\in W_1)\ i < j \Rightarrow i<'j$$
  
\noindent
The converse implication fails: $3<'6$ but $3\not< 6$.  A crisper view is the following. Linear extensions are obtained by {\it shelling a poset}, i.e. starting with any (globally) minimal element, one keeps on choosing arbitrary minimal elements of the shrinking posets. Apart from 1,2,...,8, there are many other shellings of $(W_1,\le)$, say 3,5,2,6,7,4,1,8.

\begin{center}\includegraphics[scale=0.7]{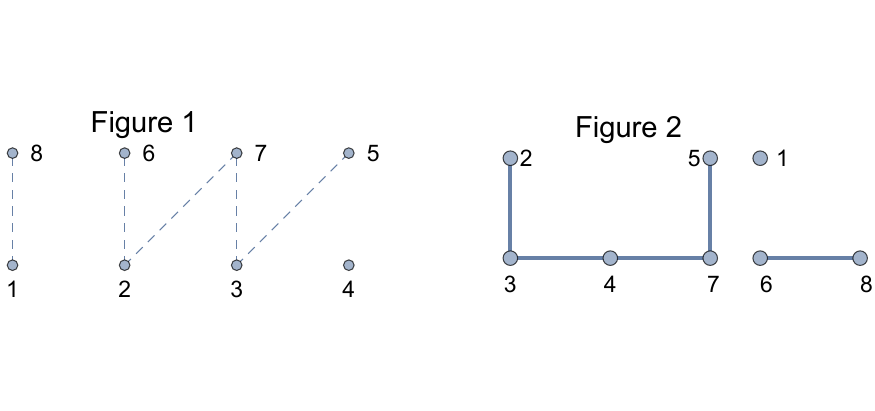}\end{center}

{\bf 2.2} Consider now the negative Boolean 2-CNF 

$$(3)\quad H_{neg,1}(x):=(\ol{x_2}\vee \ol{x_3})\wedge (\ol{x_3}\vee \ol{x_4})\wedge(\ol{x_4}\vee \ol{x_7})\wedge(\ol{x_7}\vee \ol{x_5})\wedge
(\ol{x_6}\vee \ol{x_8}).$$

\noindent
It follows that the models of $H_{neg,1}$ match the {\it anticliques} (also called independent or stable sets) of the graph $G_1$ in Figure 2.

\vspace{5mm}
{\bf 2.3} An arbitrary  clause
 is {\it Horn} if it has  at most one positive literal. By extension a {\it Horn-formula} is a conjunction of Horn clauses.
 In particular, a  Horn 2-CNF $H'(x)$ has all its clauses  of type $\ol{x_i}\vee\ol{x_j}$ or $\ol{x_i}\vee x_j$  (but not $x_i\vee x_j$ ). From $H'({\bf 0})=1$ we see that $H'$ is satisfiable. We strive for a compressed representation of $Mod(H')\neq\es$ and start with  a toy example. The systematic description of our  algorithm follows in 2.4. 
  Consider thus


\begin{itemize}
\item[(4)]    
$ H_1(x):=H_{mix,1}(x)\wedge H_{neg,1}(x)=[(\ol{x_8}\vee x_1)\wedge(\ol{x_6}\vee x_2)\wedge(\ol{x_7}\vee x_2)\wedge(\ol{x_7}\vee x_3)
\wedge(\ol{x_5}\vee x_3)]$  
\item[] $\hspace*{5cm} \wedge [(\ol{x_2}\vee \ol{x_3})\wedge (\ol{x_3}\vee \ol{x_4})\wedge(\ol{x_4}\vee \ol{x_7})\wedge(\ol{x_7}\vee \ol{x_5})\wedge
(\ol{x_6}\vee \ol{x_8}) $
\end{itemize}

\noindent
Thus $Mod(H_1)$ consists of all bitstrings $y\in \{0,1\}^8$ which simultaneously are\footnote{More precisely, the {\it support} $\{i\in [8]:\ y_i=1\}$ of $y$ simultaneously is an  ideal and an anticlique. For ease of notation we henceforth stick with $y$ and interprete it as a bitstring or as its support, at our digression.}  ideals of $(W_1,\le)$ and anticliques of $G_1$. For positive integers $k$  put $[k]:=\{1,2,...,k\}$. Let  $G_1[k]$ be the subgraph induced by $[k]$. We will calculate $Mod(H_1)$ by updating for $k=2,..,8$ the set of length $k$ bitstrings  which simultaneously are  ideals in $([k],\le)$ and anticliques in  $G_1[k]$.
Notice that $1,2,...,8$ being a shelling of $(W_1,\le)$ is crucial for $[k]$ being an  ideal of $(W_1,\le)$.

\vspace{3mm}
{\bf 2.3.1}
Glancing at Figures 1 and 2 confirms that each subset of $ \{1,2\}$ (equivalently: bitstring $y$ with $y_3=\cdots=y_8=0$) is simultaneously an  ideal of $([2],\le)$ and an anticlique of $G_1[2]$. This yields the 012-row $r_1$ in Table 1. Let us move from $k=2$ to $k=3$ and accordingly look at $r_2$ and $r_3$.

\begin{tabular}{c|c|c|c|c|c|c|c|c|l}
	& $x_1$ & $x_2$ & $x_3$ & $x_4$ & $x_5$ & $x_6$ & $x_7$ & $x_8$  & \\ \hline
	&       &       &       &       &       &       &       &              & \\ \hline
	$r_1 =$ & 2 & 2 &  &  &  &  & &  &   \\ \hline\hline

$r_2=$ & 2 & 2 &  {\bf 0} &  &0  &  &0   & & \\ \hline
$r_3=$ & 2 & 0 & {\bf 1} &0  &  &0  &0   & & \\ \hline\hline
$r_4=$ & 2 & 2 &  0 &{\bf 2}  &0  &  &0   & & \\ \hline
$r_3=$ & 2 & 0 & 1 &0  &  &0  &0   & & \\ \hline\hline
$r_5=$ & 2 & 2 &  0 &2  &0  &{\bf 0}  &0   & & \\ \hline
$r_6=$ & 2 & 1 &  0 &2  &0  &{\bf 1}  &0   &0 & final, card=4 \\ \hline
$r_3=$ & 2 & 0 & 1 &0  &  &0  &0   & & \\ \hline\hline
$r_7=$ & 2 & 2 &  0 &2  &0  &0  &0   &{\bf 0} &final, card=8 \\ \hline
$r_8=$ & 1 & 2 &  0 &2  &0  &0  &0   &{\bf 1} &final, card=4 \\ \hline
$r_3=$ & 2 & 0 & 1 &0  &  &0  &0   & & \\ \hline\hline
$r_3=$ & 2 & 0 & 1 &0  &  &0  &0   & & \\ \hline\hline
$r_9=$ & 2 & 0 & 1 &0  &{\bf 2}  &0  &0   & & \\ \hline\hline
$r_{10}=$ & 2 & 0 & 1 &0  &2  &0  &0   &{\bf 0} &final, card=4 \\ \hline
$r_{11}=$ & 1 & 0 & 1 &0  &2  &0  &0   &{\bf 1} &final, card=2 \\ \hline\hline
\end{tabular}

\vspace{3mm}
{\sl Table 1: Compressing the set of all $G_1$-independent  ideals of $(W_1,\prec)$}
\vspace{3mm}

As to $r_2$,  one can always put a $0$ at the new position $k$.
That's because an anticlique stays an anticlique in whatever way a graph increases by one vertex $k$. An  ideal stays an  ideal only because the new vertex $k$ is a {\it maximal}  element in the larger poset (due to the shelling). So much about the fat ${\bf 0}$ in $r_2$. The other 0's in $r_2$ are due to the fact that all future  ideals  $y$ with $y_3=0$ must have $y_5=y_7=0$ since $3\prec 5$ and $3\prec 7$.
As to $r_3$, one cannot always put a ${\bf 1}$ at the new position $k$. However here it  works: Since the neighborhood of $3$ in $G_1$ is $NH(3)=\{2,4\}$, and since we are aiming for anticliques, we put $y_2=y_4=0$ in $r_3$. As before, from $2\prec 6,\ 2\prec 7$ follows $y_6=y_7=0$.

The {\it pending position} to be handled in a row is the position $k$ of its first 'blank', thus $k=4$ for $r_2$. As seen, we can always fill the first blank with $0$. In $r_2$ we can also fill in 1 without altering anything else, because of $NH(k)=\{3,7\}\s zeros(r_2)$. Instead of replacing $r_2$ by two rows, one with 0, one with 1 on the blank, we
write 2 on the blank and call the new row $r_4$.
The current 
stack\footnote{This is in fact a so-called Last-In-First-Out (LIFO) stack. It is well known that LIFO stacks and depth-first-search are two sides of the same coin, see e.g. [W1, sec.7.1] for details.} consists of $r_3,\ r_4$. Always turning to the stack's top row we next handle the pending position $k=6$ of $r_4$. Filling in {\bf 0} yields $r_5$.
As to putting 1 on the 6th position, different from before 6 is no minimal element of $W_1$, and so instead of $6$ the whole 
ideal $6\!\downarrow\ =\{2,6\}$ needs to be considered. Hence $x_2=x_6=1$ in $r_6$. One has $NH(2)=\{3\}\s zeros(r_4)$, but $NH(6)=\{8\}$ forces a new 0 at position 8 in $r_6$.

 All positions of the arising row $r_6$ happen to be filled. By construction $r_6\s Mod(H_1)$, and so $r_6$ is {\it final}. It is removed from the  stack and stored in a safe place. One verifies that handling position 8 of the new top row $r_5$ yields rows $r_7$ and $r_8$. Both of them are final, and so only $r_3$ remains in the  stack.
It holds that $5\!\downarrow\ =\{3,5\}$ and $3\in ones(r_3)$; further $NH(5)=\{7\}\s zeros(r_3)$. Hence the blank on the 5th position of $r_3$ can be filled with 2, giving rise to row $r_9$. In turn $r_9$ gives rise to the final rows $r_{10}$ and $r_{11}$. It follows that there are $|r_6|+|r_7|+|r_8|+|r_{10}|+|r_{11}|=22$  ideals of $(W_1,\preceq)$ which are $G_1$-independent.

\vspace{5mm}
{\bf 2.4} For any poset $(P,\le)$ and any $a\in P$ we adopt the notations $a\!\downarrow\ :=\{b\in P:\ b\le a\}$ and $a\!\uparrow\ :=\{b\in P:\ b\ge a\}$. More generally
$Z\!\downarrow:=\bigcup \{a\!\downarrow\ :\ a\in Z\}$, and likewise $Z\!\downarrow$. 

\vspace{2mm}
Let $(W,\le)$ and $G$ be a poset and graph respectively that share the same vertex-set $W=[w]$. We may assume that $1,2,...,w$ is a shelling of $(W,\le)$.
 Let us state systematically how to extend a partial row $r$ to $r'$ and $r''$ by filling its first blank (at position $k$) by 0 and 1 respectively. Inducting on $k=1,2,..,w$ these properties need to be maintained:

\begin{itemize}
	\item[(P1)] Up to position $k-1$ there are no blanks\footnote{It helps to  distinguish (also notationally) the preliminary 2's up to position $k-1$ from the other preliminary 2's. The latter we call {\it blanks}. They match the (ordinary) blanks in Table 1.}, and if $k\le i\le w$ then the $i$th position is either $0$ or a blank.
	\item[(P2)] Whenever $i\in zeros(r)$, then $(i\!\uparrow)\s zeros(r)$.
	\item[(P3)]  Whenever $i\in ones(r)$, then $(i\!\downarrow)\s ones(r)$.
	\item[(P4)] The  set $ones(r)$ is $G$-independent.	
	\end{itemize}

Here comes the recipe to maintain these properties when moving from $r$ to $r'$ and $r''$:


\begin{itemize}
	\item[(R1)] The  set $k\!\uparrow$ is a subset of $\{k,k+1,..,w\}$ in view of the shelling order. Hence by (P1) for each $i\in k\!\uparrow$ the $i$th position in $r$ is either already $0$ or a blank, and so one can write a $0$ on it. It is clear that (P1) to (P4) are maintained by the new partial row $r'$.
	
	\item[(R2)] By the shelling order  the position set $k\!\downarrow$ is a subset of $[k]=\{1,2,..,k\}$. By (P2) it holds that $(k\!\downarrow)\cap zeros(r)=\emptyset$. Hence one can write 1 on the $i$th position for all $i\in k\!\downarrow$, {\it provided} (Case 1) that $Y:=(k\!\downarrow)\cup ones(r)$ happens to be $G$-independent. In the latter case define $Z\s [w]$
	as the set of $j$ that are adjacent (in $G$) to some $i\in Y$. Then $Z\cap ones(r)=\emptyset$ since $Y$ is $G$-independent.Hence each component of $r$ whose index $i$ is in $Z$, is either $0$ or $2$. Using (P2) and (P3) respectively it follows that $i\!\uparrow\cap ones(r)=\emptyset$. Therefore  $(Z\!\uparrow)\cap ones(r)=\emptyset$, and so we can write 0 on all positions $i\in Z\!\uparrow$. One checks that (P1) to (P4) are maintained by the extension $r''$ of $r$. 
    
    If (Case 2) $Y$ is $G$-dependent, then only $r'$, not $r''$, is built.
	
	\item[(R3)] Let $r'$ and $r''$ be the new partial rows arising in (R1) and (R2) respectively. Suppose $zeros(r')$ is as small as it can possibly be, i.e. $zeros(r')=zeros(r)\cup\{k\}$. Likewise assume that  $ones(r'')=ones(r)\cup\{k\}$ and $zeros(r'')=zeros(r)$. Then, the two rows $r',r''$ can be replaced by a single row $r'''$ that arises from $r$ by writing '2' on the first blank.
		
\end{itemize}

{\bf Theorem 1:} {\it Let $(W,\preceq)$ be a $w$-element poset and $G$ a graph with vertex set $W$. Then the above algorithm {\tt All-Independent-Ideals (A-I-I)} represents the set of all $G$-independent  ideals as a disjoint union of $R$ many 012-rows in time $O(Rw^3)$.}

\vspace{4mm}
{\it Proof.} By the workings of A-I-I layed out above it suffices to show that (correctly) filling in a blank  costs $O(w^2)$, and that this happens $O(Rw)$ many times. As to the first claim, ckecking the independency of $(k\!\downarrow)\cup ones(r)$ in (R2) costs $O(w^2)$, and this swallows all other costs (such as calculating $k\!\downarrow$ and $Z$). As to the second claim, let $A$ be the set of the $Rw$ many  components occuring in the $R$ many final rows; and let $B$ the set of all 'blank-filling events'. Then there is an obvious (well-defined) function $f$ from $A$ to $B$. Crucially, $f$ is surjective\footnote{But unless there is just one final row, $f$ is not injective. To spell it out, suppose the  partial row $r$ splits, and $c$ is the entry to the left of the blank that triggers the splitting. Then the blank-filling event that produced $c$ is the $f$-value of two {\it distinct} elements of $A$. } since partial rows are never deleted. This proves the second claim. $\square$
	
The algorithm A-I-I will be (informally) extended in Sections 4 and 5. Both times the dedicated  reader will have no problems formulating a concise statement and verifying that the bound $O(Rw^3)$ remains the same.

\section{Numerical experiments}

We compare our high-end Mathematica\footnote{Specifically, I used Mathematica 14.2 on a Dell Latitude 7420 laptop, with an 11-th generation Intel i7  CPU, and 32 GB memory.} implementation of {\tt All-Independent-Ideals} (A-I-I) with the Mathematica command {\tt BooleanConvert}  (option 'ESOP') which converts any Boolean function $F$ in a so-called 'exclusive sum of products'.  In effect $Mod(F)$ gets written as a disjoint union of 012-rows. Since the Mathematica command {\tt SatisfiabilityCount} (={\tt SCount} in Table 2) merely calculates\footnote{Note that {\tt SatisfiabilityCount} uses BDD's as explained in 7.2. As to the theoretic complexity to count the models of a 2-CNF, see [FK].} the cardinality $|Mod(F)|$, it has an inherent advantage over A-I-I and {\tt BooleanConvert} (={\tt BConvert} in Table 2). Being "hard-wired" Mathematica commands, both {\tt BooleanConvert} and  {\tt SatisfiabilityCount} have a headstart on A-I-I.

\vspace{4mm}
{\bf 3.1} Specifically, similarly to [W2, p.132] we generate random posets $(W,\le)$ which consist of $he+1$ many levels $Lev(i)$, all of cardinality $br$, in such a way that each $a\in Lev(i)\ (2\le i\le he+1)$ is assigned $lc$ random lower covers in $Lev(i-1)$. Thus $(W,\preceq)$ has breadth $br$, height $he$ and cardinality $w:=|W|=br(he+1)$. (For instance, except for the non-uniform value of $lc$, Figure 1 depicts such a poset with $br=4$ and $he=1$.)

Furthermore a graph $G$ with the same vertex set $W$ and $m$ random edges is created. Then A-I-I computes all $G$-independent  ideals as described in 2.3 and 2.4. 
(To check $G$-independency in (R2) we used the Mathematica command {\tt IndependentVertexSetQ}.) The number of models\footnote{It goes without saying (almost) that all three algorithms, albeit extremely different, always convened on the same number of models. Verifying the correctness of {\tt BooleanConvert} often took longer than the task itself; e.g; applying 1’798’229 times {\tt SatisfiabilityCount} and adding up the values would have taken longer than the time for A-I-I.} is recorded in the column labelled '$\#$ models'. The various CPU-times in sec are recorded in the columns labelled A-I-I,  {\tt SCount} and {\tt BConvert}. Finally the number of final 012-rows produced by A-I-I and {\tt BooleanConvert} are recorded in the columns labelled $R_{AII}$ and $R_{BC}$ respectively.

 \vspace{3mm}
 
 \begin{tabular}{c|c||c||c|c||c|c||l}
 	 $(br,he,lc)\ra (w,m)$ & $\#$ models & {\tt SCount} & {\tt BConvert} & $R_{BC}$  & A-I-I & $R_{AII}$   &  \\ \hline
 	       &       &       &       &       &       &       &               \\ \hline
 	  $(15,4,2)\ra (75,20)$ & 7'774'472 & 0.4 s &  0.3 s & 12'456  & 4.7 s  &53'264   &  \\ \hline
 	   $(15,4,2)\ra (75,1000)$ &469 & 0.7 s &  0.39 s & 68  & 0.05 s  &142   &  \\ \hline
 	    $(30,6,8)\ra (210,7000)$ & 8150 & 717 s &  923 s & 505  & 3.4 s  &2566   &  \\ \hline\hline

 $(40,1,10)\ra (80,15)$ & $\approx 853\cdot 10^9$ & 238 s 
           &  137 s & 1'798'229  & 1268 s  & 7'917'456   &  \\ \hline
  $(40,1,10)\ra (80,20)$ & $\approx 475\cdot 10^9$ & 245 s 
           &  131 s & 185'253  & 105 s  & 640'586 &  \\ \hline
 $(40,1,10)\ra (80,100)$ & $\approx 8\cdot 10^9$ & 210 s 
           &  61 s & 10'420  & 20 s & 108'136   &  \\ \hline
           
 $(40,1,10)\ra (80,600)$ & $554'784$ & 196 s 
           &  95 s & 9449  & 11 s  & 46'760   &  \\ \hline
 $(40,1,10)\ra (80,2000)$ & $1138$ & 154 s 
           &  58 s & 286  & 0.2 s  & 546   &  \\ \hline\hline

 	    $(42,1,10)\ra (84,700)$ & 841'973 & 427 s &  111 s & 8767  & 19 s  &61'621   &  \\ \hline
 	    $(44,1,10)\ra (88,800)$ & 785'469 & aborted &  362 s & 16'205  & 29 s  &111'289   &  \\ \hline
 	    $(48,1,10)\ra (96,1100)$ & 1'558'461 & aborted & aborted & ---  & 61 s  &215'680   &  \\ \hline
         $(200,1,10)\ra (400,40000)$ & 6'563'864 & aborted &  aborted & ---  & 18'708 s  &2'421'566   &  \\ \hline
 	       $(200,1,10)\ra (400,46000)$ & 1'151'586 & aborted &  aborted & ---  & 2689 s  &515'895   &  \\ \hline
            $(200,1,10)\ra (400,50000)$ & 456'495 & aborted &  aborted & ---  & 1729 s  &221'599   &  \\ \hline\hline

             $(3,27,1)\ra (84,0)$ & $\approx 234\cdot 10^9$ & 0 s &  0.5 s & 76484 & 207 s  &1'151'622   &  \\ \hline\hline

 \end{tabular}

 \vspace{3mm}
{\sl Table 2. Computational experiments with A-I-I, {\tt BooleanConvert}, and  {\tt SatisfiabilityCount}.}

 \vspace{3mm}

  Note that always $R_{BC}<R_{AII}$, thus  compression-wise A-I-I trails\footnote{We mention in passing that the situation is often reversed when wildcards beyond don't-cares are pitted against {\tt BooleanConvert}, e.g. see [arXiv.1812.02570v3].}   {\tt BooleanConvert}. Concerning CPU-times, either one of the two can be the winner. 

\vspace{4mm}
{\bf 3.2}
 Specifically, consider the instances  $(40,1,10)\ra (80,m)$. While ${\tt BooleanConvert}$ prevails for $m=15$, as $m$ increases (and hence the 2-CNF) the times for A-I-I fall more drastically than for ${\tt BooleanConvert}$. Let us look at the instances $(br,1,10)\ra (*,*)$, where $(*,*)$ is tuned to keep the number of models below 10'000'000. 
 As $br$ increases, the time for A-I-I increases more benign (to put it mildly) than the times of {\tt BooleanConvert}, let alone  {\tt SatisfiabilityCount}.
Both repeatedly  self-aborted within half an hour due to memory problems. Even with infinite memory, extrapolating their  CPU-times for  $br=40,42,44$ raises the question whether $br=200$ could be handled in this century. 
As to the last instance $(3,27,1)\ra (84,0)$, see the remarks in 6.1.

\vspace{3mm}
 {\bf 3.3} Even when A-I-I cannot finish within reasonable time, it can (like every LIFO-based algorithm, but different from {\tt BooleanConvert}) deliver results as it goes on. Thus an instance of type $(42,22,7)$ was stopped after 24 hours, having produced 27'707'488 rows comprising 323'242'512'184 models. The reader may ask: Why would one wish to produce that many models? True, but an idea of their "average properties"  would be nice. In this context observe that a little extra effort (see [W3, Sec.6.3)]) yields an  {\it uniform random sample} of desired cardinality of the total modelset.
 Furthermore, A-I-I is easy\footnote{As to {\tt BooleanConvert}, only its programmer(s) can tell.} to parallelize, as is {\it every} algorithm based on  a LIFO stack (as to why, see e.g [W3,Sec.6.5]).
 
 Section 5 will show how any satisfiable 2-CNF $F$ yields 
an "equivalent" Horn 2-CNF $H'$ in linear time. Section 4 below shows how $H'$ can (again in linear time) be turned to an acyclic formula $H$ which hence can be digested by A-I-I.

\section{Making Horn 2-CNFs fit for the algorithm A-I-I}

Here we transform an arbitrary Horn 2-CNF $H'$ to some Horn 2-CNF $H$ that can be handled by A-I-I. Since it will get a bit technical, keep in mind: It all boils down to factor out the strong components of a digraph, which is long known to be doable in linear time.

In the sequel we usually write $x_i\to x_j$ for each mixed clause $\ol{x}_i\vee x_j$. Consequently the mixed clauses of each Horn 2-CNF $H'$ define a (loopless) digraph  $(W,\to)$ on the set $W:=\{x_1,...,x_w\}$ of Boolean variables. If there are {\it no directed cycles} in $(W,\to)$, then the reflexive-transitive closure $\ge$ of the binary relation $\to$  yields a  poset $(W,\ge)$. For $H_{mix,1}$ in (1) we had $(W_1,>)=(W_1,\to)$, recall Fig.1.

\vspace{5mm}
{\bf 4.1} Let us step up $H_{mix,1}$  by looking at

\begin{itemize}

\item[(5)] $H_{mix,2}:=(x_3\to x_1)\wedge (x_3\to x_2)\wedge(x_4\to x_2)\wedge(x_5\to x_2)\wedge(x_5\to x_3)$

\item[] $\hspace{1.4cm} \wedge(x_5\to x_4)\wedge(x_6\to x_3)\wedge(x_6\to x_4)
\wedge(x_7\to x_5)\wedge(x_7\to x_6)$
\end{itemize}

\noindent
If $W_2:=\{x_1,...,x_7\}$ and $\to$ comes from (5), then the digraph $(W_2,\to)$ is without directed cycles, and hence yields a poset $(W_2,\ge)$.
For instance $x_7>x_1$ is a consequence of $(x_7\to x_6)\wedge(x_6\to x_3)\wedge(x_3\to x_1)$. By convention the {\it diagram} [CLM,p.7]  of a poset only features the {\it covering} pairs $x\succ y$ (so $\not\exists z$ with $x>z>y$). 
The diagram of the poset $(W_2,\ge)$ derived from $H_{mix,2}$ is rendered in Fig.3i. As in the general case, all coverings of $(W_2,\ge)$ are among\footnote{Note that $x_5\to x_2$ from (5) yields $x_5>x_2$, but this is no covering.} the arrows $x_i\to x_j$ appearing (5).

\vspace{2mm}
We now blow up $H_{mix,2}$ to

\begin{itemize}
    \item[(6)] $H_{mix,2}':=H_{mix,2}\wedge(x_1\to x_{11})\wedge(x_{11}\to x_1)\wedge(x_4\to x_{12})\wedge(x_{12}\to x_{13})\wedge(x_{13}\to x_4)$
    \item[] $\hspace{2.2cm}\wedge(x_6\to x_8)\wedge(x_8\to x_9)\wedge(x_9\to x_6)\wedge(x_6\to x_{10})\wedge(x_{10}\to x_9). $
\end{itemize}

\noindent
Now the induced digraph $D(H_{mix,2}')$ has four directed cycles (see Fig.3ii). 
One gets  a rough idea of $D(H_{mix,2}')$ as a whole  by looking at Fig.3i and Fig.3ii simultaneously.
The usual procedure to simplify digraphs is to put $x_i\sim x_j$ iff there is a directed path from $x_i$ to $x_j$, and one from $x_j$ to $x_i$.
This is an equivalence relation whose classes, called the {\it strong components}, constitute the vertices of the {\it factor-poset} derived from the digraph. In our case the {\it proper} ($\ne$ singleton) strong components of $D:=D(H_{mix,2}')$ are $\{x_1,x_{11}\},\{x_4,x_{12},x_{13}\},\{x_6,x_8,x_9,x_{10}\}$. The factor-poset $\widetilde{D}$ of $D$ happens to be isomorphic to the poset in Fig.3i.

\begin{center}\includegraphics[scale=0.7]{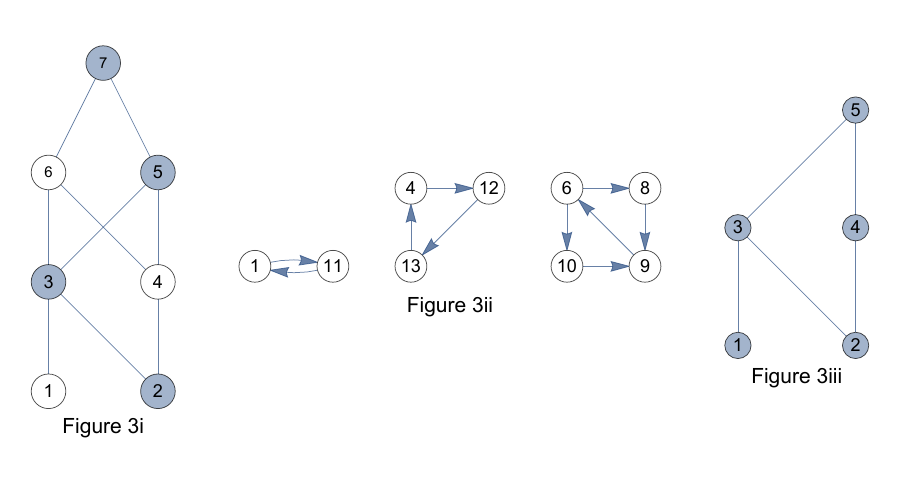}\end{center}

\vspace{3mm}
 {\bf 4.1.1} However in our scenario it doesn't stop with the factor-poset. Namely, suppose $y\in \{0,1\}^{13}$ is any model of $H'_{mix,2}$. It is easy to see  that $y$ is constant on each strong component; thus for instance either $y_6=y_8=y_9=y_{10}=0$ or
 $y_6=y_8=y_9=y_{10}=1$. Unfortunately, being constant may lead to conflicts if negative clauses are present. To fix ideas, consider\footnote{As to using the dash $'$, we write $H_{mix}'$ if the structure of the induced digraph is unknown or irrelevant. If the latter is acyclic (i.e. all strong components are singletons), we may write $H_{mix}$. As to $H'_{neg}$ versus $H_{neg}$ we may use whatever looks more pleasant (What concerns negative clauses on their own, there are no strong components to worry about.)}

 \begin{itemize}
     \item[(7)] $H'_{neg,2}:=(\ol{x}_1\vee\ol{x}_4)\wedge (\ol{x}_1\vee\ol{x}_{12})\wedge(\ol{x}_2\vee\ol{x}_7)\wedge(\ol{x}_3
     \vee\ol{x}_5)\wedge(\ol{x}_4\vee\ol{x}_5)\wedge(\ol{x}_4\vee\ol{x}_9)$
     
     \item[]$\hspace{1.4cm}\wedge(\ol{x}_5
     \vee\ol{x}_6)\wedge(\ol{x}_5\vee\ol{x}_7)\wedge(\ol{x}_6\vee\ol{x}_8)\wedge (\ol{x}_9\vee\ol{x}_{10})\wedge (\ol{x}_{10}\vee\ol{x}_{13})\wedge(\ol{x}_{11}\vee\ol{x}_{13}).$
 \end{itemize}

\noindent
The graph $G(H'_{neg,2})$ whose anticliques match the models of $H'_{neg,2}$ is rendered in Fig.4i. Notice that the strong component $\{x_6,x_8,x_9,x_{10}\}$ of $D(H_{mix,2}')$ is  "bad" in the sense that no model $y\in Mod(H'_{neg,2}$ satisfies $y_6=y_8=y_9=y_{10}=1$ because this contradicts $\ol{y}_6\vee\ol{y}_8=1$
(and also $\ol{y}_9\vee\ol{y}_{10}=1$). In other words, each $y\in Mod(H'_{neg,2})$ satisfies $y_6=y_8=y_9=y_{10}=0$.

\vspace{3mm}
We put $H_2':=H_{mix,2}'\wedge H'_{neg,2}$ and
$D(H_2'):=D(H_{mix,2}')$ as well as $G(H_2'):=G(H'_{neg,2})$. 
So far, so good. However, the digraph $D(H_2')$ (Fig.3i,3ii) and the graph $ G(H_2')$ (Fig.4i) will affect each other to the extent that both shrink considerably. That is, $D(H_2')$ becomes the poset $P(H_2')$
in Fig.3iii, and $G(H_2')$ becomes the small graph $G^*(H_2')$ in Fig.4iii.  Details follow in 4.2.

\begin{center}\includegraphics[scale=0.96]{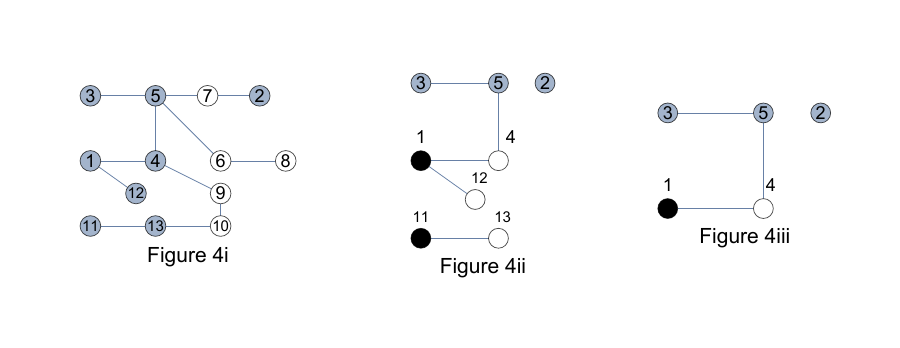}\end{center}

{\bf 4.2} Let $H'$ be any Horn 2-CNF with set of variables $W$. The corresponding graph $G(H')$ and digraph $D(H')$ are defined analogous to $G(H_2')$ and $D(H_2')$. 
In 4.2.1 we will shrink $D(H')$  to some poset $P(H')$, and in 4.2.2 shrink $G(H')$  to some  graph $G^*(H')$. It will help to illustrate these manipulations for the concrete case $H'=H_2'$.

\vspace{3mm}

{\bf 4.2.1}  A strong component $S\s W$ of $D(H')$ is {\it bad} iff $S$ is {\it no} anticlique in the coupled graph $G(H')$ on the same set $W$. If $S_1,...,S_t$ are the bad strong components (thus elements of the factor-poset $\widetilde{D}(H')$, then removing the {\it doomed\footnote{Observe that not all {\it members} of $Fi$ need be bad; e.g. when $Fi$ belongs to $H_2'$, then $\{7\}\in Fi$ is not a bad strong component (since $\{7\}$ is an anticlique of $G(H_2')$). Nevertheless $7$ needs to be removed as well because $y_7=0$ for all $y\in Mod(H_2')$ (why?).} filter } $Fi:=\{S_1,..,S_t\}\uparrow$ yields the {\it base poset} $P(H')$; see Fig.3iii for $P(H_2')$.

\vspace{3mm}

{\bf 4.2.2} In  order to obtain an analogous {\it base graph} $G^*(H')$ from $G(H')$ first remove all vertices in $Fi$ (and the edges incident with them) from  $G(H')$. Afterwards the vertices in each remaining proper strong component are condensed to a single vertex which however retains the neighbors of its condensed forfathers.

To fix ideas, if $H'=H_2'$, then $\bigcup Fi=\bigcup\{\{6,8,9,10\},\{7\}\}=\{6,7,8,9,10\}$. The graph obtained from $G(H_2')$ upon removing $Fi$ is rendered in Figure 4ii. In this graph we condense the proper strong components $\{1,11\}$ and $\{4,12,13\}$ and obtain $G^*(H_2')$ in Figure 4iii.

\vspace{4mm}
{\bf 4.3} The reader may verify that running {\tt All-Independent-Ideals} on the base poset $P(H_2')$ and the base graph $G^*(H_2')$ (with common universe $\{1,2,..,5\}$) yields the final rows

\begin{center}
$\rho_1=(1,1,1,0,0)$ and $\rho_2=(2,2,0,0,0)$ and $\rho_3=(0,1,0,1,0)$. \end{center}

How to get $Mod(H_2')\s\{0,1\}^{13}$ from this? First, all variables having indices in $\bigcup Fi=\{6,7,8,9,10\}$ are set to zero. While each $y\in Mod(H_2')$ is doomed to be $0$ on the strong component $\{6,8,9,10\}$, on the other strong components
$\{1,11\}$ and $\{4,12,13\}$ our $y$ only needs to be {\it constant}.
 Hence if $y_1=\underline{1}$  then\footnote{We write $\underline{1}$ instead of $1$ for visual effect in Table 3.} $y_{11}=\underline{1}$, and if $y_1=\underline{0}$  then $y_{11}=\underline{0}$. Similarly the value of $y_4$ is adopted by $y_{12},y_{13}$. This explains why $\ol{\rho}_1,\ol{\rho}_3$ in Table 3 belong to $Mod(H_2')$.

As to $\rho_2$, upon splitting it into $\rho_{2,0}\uplus\rho_{2,1}:=(0,2,0,0,0)\uplus (1,2,0,0,0)$ it similarly follows that $\ol{\rho}_{2,0}$ and $\ol{\rho}_{2,1}$ in Table 3 belong to $Mod(H_2')$.

\vspace{3mm}

\begin{tabular}{c|c|c|c|c|c||c|c|c|c|c||c|c|c|}
& 1 & 2& 3 &4  &5 &6 &7 &8 &9 &10 &11 &12 &13\\ \hline
$\ol{\rho}_1$ & $\underline{1}$ & 1&1&{\bf 0}&0& 0&0&0&0&0   &$\underline{1}$&{\bf 0}&{\bf 0}\\ \hline
$\ol{\rho}_{2,0}$ & $\underline{0}$& 2&0&{\bf 0}&0& 0&0&0&0&0   &$\underline{0}$&{\bf 0}&{\bf 0}\\ \hline
$\ol{\rho}_{2,1}$ & $\underline{1}$& 2&0&{\bf 0}&0& 0&0&0&0&0   &$\underline{1}$&{\bf 0}&{\bf 0}\\ \hline
$\ol{\rho}_3$ & $\underline{0}$& 1&0&{\bf 1}&0& 0&0&0&0&0   &$\underline{0}$&{\bf 1}&{\bf 1}\\ \hline
\end{tabular}

\vspace{3mm}
{\sl Table 3: The modelset of the Horn 2-CNF $H_2'$  from 4.1.1. }

\vspace{3mm}
{\bf 4.4} The first sentence in Section 4 promised to reduce any Horn 2-CNF $H'$ to  another Horn 2-CNF $H$ which can be digested by A-I-I. So, given $H_2'$, where is $H_2$?
With hindsight we see that spelling out $H_2$ was not necessary; all that matters is to know the base poset $P(H_2')$ and the base graph $G^*(H_2')$. But if forced, it is easy to spell out $H_2$; here is one possibility: 

\begin{center}
    $H_2:=(x_5\to x_3)\wedge (x_5\to x_4)\wedge (x_3\to x_1)\wedge(\ol{x}_1\wedge\ol{x}_4)\wedge(\ol{x}_4\wedge\ol{x}_5)\wedge(\ol{x}_5\wedge\ol{x}_3)$.
    \end{center}
    
Consequently $Mod(H_2)=\rho_1\uplus\rho_2\uplus\rho_3$. While the way from $H'$ to $H$ may seem tedious, computationally it works very fast.

\section{Reducing arbitrary  2-CNFs to Horn 2-CNFs}

In the introduction we claimed that "switching some magic set of variables" of a satisfiable 2-CNF $F$ yields a {\it Horn} 2-CNF $H'$. (Recall, upon turning $H'$ to an {\it acyclic} formula $H$ as shown in Sec. 4, we can apply A-I-I to $H$.) We prove the claim in 5.1 and follow up with a toy example in 5.2. In 5.3 we investigate under what circumstances we can get extra nice Horn 2-CNFs.

\vspace{3mm}
{\bf 5.1} According to  [S, Thm. 5.19] the following holds. 
 Suppose a Boolean 2-CNF $F(x)$ is satisfiable, and
$y$ is any fixed model. If $H(x)$ is the Boolean function obtained from $F(x)$ by switching those variables $x_i$ for which $y_i=1$, then $H(x)$ is a Horn formula. Thus satisfiable 2-CNFs are {\it Horn-renamable}. In [S] the proof of Thm. 5.19 is left as exercise. Let us give an outline. 

So why, upon switching the indicated variables, does each  $F$-clause become (or stay) either negative or mixed? Starting with negative clauses $\ol{x_i}\vee\ol{x_j}$, the only way for them to turn  positive is by switching both literals. This happens iff $y_i=y_j=1$, i.e. iff $\ol{y_i}\vee\ol{y_j}=0$. But then $F(y)=0$, contradicting the assumption that $F(y)=1$. Similar reasoning shows that each positive $F$-clause becomes mixed or negative, and that each mixed $F$-clause stays mixed of becomes negative.

\vspace{3mm}
{\bf 5.2} Consider the Boolean 2-CNF $F_2(x)$ whose clauses are displayed in the left column of Table 4. Because of the several positive clauses it is far from being Horn. 
Define $y\in \{0,1\}^{13}$ by $y_6=y_{11}=y_{12}:=1$, and $y_i:=0$ otherwise. One checks that $y\in Mod(F_2)$. Hence, if we accordingly switch the variables $x_6,x_{11},x_{12}$, then $F_2$ becomes some Horn formula $H'$. In fact, incidently $H'=H_2'$ from above\footnote{The top/middle/bottom part of the second column in Table 3 match equations (5),(6),(7).}. Knowing $Mod(H_2')$ (see Table 3) we conclude that $Mod(F_2)=\sigma_1\uplus\sigma_{2,0}\uplus\sigma_{2,1}\uplus \sigma_3$, where each $\sigma$-row arises from its equally indexed companion in Table 3 by switching $0$ with $1$, whenever either  is in position 6,11 or 12. For instance $\sigma_1=(1,1,1,0,0,{\bf 1},0,0,0,0,{\bf 0,1,}0)$.

\vspace{3mm}
\begin{tabular}{c|c|c}
    $F_2(x)$ & $H_2(x)$ & change of clause? \\ \hline\hline
    $\ol{x}_3\vee x_1$ &  $\ol{x}_3\vee x_1$ &  \\ \hline  
     $\ol{x}_3\vee x_2$ &  $\ol{x}_3\vee x_2$& \\ \hline
 $\ol{x}_4\vee x_2$ &  $\ol{x}_4\vee x_2$& \\ \hline
  $\ol{x}_5\vee x_2$ &  $\ol{x}_5\vee x_2$& \\ \hline
   $\ol{x}_5\vee x_3$ &  $\ol{x}_5\vee x_3$& \\ \hline
    $\ol{x}_5\vee x_4$ &  $\ol{x}_5\vee x_4$& \\ \hline
     $pos:\ x_6\vee x_3$ &  $\ol{x}_6\vee x_3$& yes \\ \hline
      $pos:\ x_6\vee x_4$ &  $\ol{x}_6\vee x_4$& yes \\ \hline
       $\ol{x}_7\vee x_5$ &  $\ol{x}_7\vee x_5$& \\ \hline
  $\ol{x}_7\vee \ol{x}_6$ &  $\ol{x}_7\vee x_6$& yes \\ \hline\hline

 $\ol{x}_1\vee \ol{x}_{11}$ &  $\ol{x}_1\vee x_{11}$& yes \\ \hline
 $pos:\ x_{11}\vee x_1$ &  $\ol{x}_{11}\vee x_1$& yes \\ \hline
 $\ol{x}_4\vee \ol{x}_{12}$ &  $\ol{x}_4\vee x_{12}$& yes \\ \hline
 $pos:\ x_{12}\vee x_{13}$ &  $\ol{x}_{12}\vee x_{13}$& yes \\ \hline
 $\ol{x}_{13}\vee x_4$ &  $\ol{x}_{13}\vee x_{4}$& \\ \hline
 $pos:\ x_6\vee x_8$ &  $\ol{x}_6\vee x_{8}$& yes \\ \hline
 $\ol{x}_8\vee x_{9}$ &  $\ol{x}_8\vee x_{9}$& \\ \hline
 $\ol{x}_9\vee \ol{x}_{6}$ &  $\ol{x}_9\vee x_{6}$& yes \\ \hline
 $pos:\ x_6\vee x_{10}$ &  $\ol{x}_6\vee x_{10}$& yes \\ \hline
 $\ol{x}_{10}\vee x_{9}$ &  $\ol{x}_{10}\vee x_{9}$& \\ \hline\hline

 $\ol{x}_1\vee \ol{x}_{4}$ &  $\ol{x}_1\vee \ol{x}_{4}$& \\ \hline
 $\ol{x}_1\vee x_{12}$ &  $\ol{x}_1\vee \ol{x}_{12}$& yes \\ \hline
 $\ol{x}_2\vee \ol{x}_{7}$ &  $\ol{x}_2\vee \ol{x}_{7}$& \\ \hline
 $\ol{x}_3\vee \ol{x}_{5}$ &  $\ol{x}_3\vee \ol{x}_{5}$& \\ \hline
 $\ol{x}_4\vee \ol{x}_{5}$ &  $\ol{x}_4\vee \ol{x}_{5}$& \\ \hline
 $\ol{x}_4\vee \ol{x}_{9}$ &  $\ol{x}_4\vee \ol{x}_{9}$& \\ \hline
 $\ol{x}_5\vee x_6$ &  $\ol{x}_5\vee \ol{x}_{6}$& yes \\ \hline
 $\ol{x}_5\vee \ol{x}_{7}$ &  $\ol{x}_5\vee \ol{x}_{7}$& \\ \hline
 $x_6\vee \ol{x}_{8}$ &  $\ol{x}_6\vee \ol{x}_{8}$& yes \\ \hline
 $\ol{x}_9\vee \ol{x}_{10}$ &  $\ol{x}_9\vee \ol{x}_{10}$& \\ \hline
 $\ol{x}_{10}\vee \ol{x}_{13}$ &  $\ol{x}_{10}\vee \ol{x}_{13}$& \\ \hline
 $x_{11}\vee \ol{x}_{13}$ &  $\ol{x}_{11}\vee \ol{x}_{13}$& yes \\ \hline\hline  
\end{tabular}

\vspace{3mm}
{\sl Table 4: $F_2(x)$ becomes a Horn formula upon switching $x_6,x_{11},x_{12}$}

\vspace{4mm}
{\bf 5.3} We saw in 5.1 and 5.2 that the satisfiability of a 2-CNF $F$ is sufficient for being Horn-renamable. The obtained Horn 2-CNF $H'$ will be  even more useful if its clauses happen to be all mixed (so $H'$ is of type $H_{mix}'$), or they all are negative ($H'=H'_{neg}$). Can one influence "happen to" by fine-tuning the Horn-renaming? The answer is yes, both for getting negative (5.3.1) and mixed (5.3.2) clauses.

\vspace{3mm}
{\bf 5.3.1} In order to transform an arbitrary 2-CNF $F$ to type $H'_{neg}$ we look at the three types of clauses $C$ that $F$ may have. If $C=x_i\vee x_j$ then we must switch both $x_i$ and 
$x_j$. If $C=x_i\vee\ol{x}_j$ then only $x_i$ must be switched, and if $C=\ol{x}_i\vee\ol{x}_j$ then none of the variables must be switched. Thus no choice  is possible. (We leave it to the reader to devise instances of both success and failure.)

\vspace{3mm}
{\bf 5.3.2} Deciding whether an arbitrary 2-CNF $F(x)$ can be "switched" to type $H'_{mix}$ is more subtle. We set up an auxiliary 2-CNF $Aux(s)$ whose  variables $s_i$ bijectively match the variables $x_i$ of $F(x)$. Furthermore the 2-clauses of $Aux(s)$ arise as follows.
Again we look at the three types of clauses $C$ that $F$ may have. If $C=x_i\vee x_j$ then exactly one variable must be switched. This amounts\footnote{In terms of clauses this is
 $(s_i\vee s_j)\wedge(\ol{s}_i\vee\ol{s}_j)=1$.} to $s_i\oplus s_j=1$ (addition modulo 2). Likewise, if $C=\ol{x}_i\vee\ol{x}_j$, then $s_i\oplus s_j=1$. And if $C=\ol{x}_i\vee x_j$, then $s_i\oplus s_j=0$.

 We see that obtaining $H_{mix}'$ is possible iff some system of linear equations over the 2-element field is solvable. If the system is inconsistent, then solving as many equations as possible, will still be beneficial later on.

 \section{Wildcards beyond the don't-care symbol}

 We ponder under what circumstances {\tt All-Independent-Ideals} may be favorably replaced by  thoroughly different methods that employ wildcards beyond the don't-care symbol.
Let us disclose the bottom line right away, before being distracted  by details:
Yes,  better compression is possible for the three types of {\it homogeneous} 2-CNFs. For other scenarios there still is potential for compression. But even for acyclic Horn 2-CNFs (which are taylor-made for A-I-I) achieving extra compression will be hard.

 \vspace{5mm}
 {\bf 6.1} Let $H_{mix}$ be any {\it acyclic} Horn formula with exclusively mixed clauses, and let $H_{neg}$ be any Horn formula with exclusively negative clauses.
 In the past (before discovering A-I-I) the author has tackled the enumeration of $Mod(H_{mix})$ and $Mod(H_{neg})$. Let us briefly survey the two methods in 6.1.1 and 6.1.2.
 (As for $Mod(H_{pos})$, see 6.4.1.)

 \vspace{5mm}
 {\bf 6.1.1} The observation that (say) $(x_1\to x_2)\wedge(x_1\to x_3)\wedge(x_1\to x_4)$ is equivalent to $x_1\to(x_2\wedge x_3\wedge x_4)$, and that $Mod(x_1\to(x_2\wedge x_3\wedge x_4))=(1,1,1,1)\uplus(0,2,2,2)$, gives rise to the {\it (a,b)-wildcard}. Thus by definition $(a,b,b,b):=(1,1,1,1)\uplus(0,2,2,2)$. More generally [W2], a typical {\it 012ab-row} would be $r:=(a_1,2,b_1,a_3,b_2,1,a_2,b_3,0,b_3)$,
  where $a_1b_1,\ a_2b_2,\ a_3b_3b_3$ (dispersed throughout $r$) are mutually independent (a,b)-wildcards. The gain in compression is considerable because it would take eight 012-rows to represent $Mod(r)$. Specifically, by substituting 0,1 to $a_1,a_2,a_3$ in all possible ways one obtains eight 012-rows whose disjoint union is $Mod(r)$. A random one of these 8 rows is $({\bf 0},2,2,{\bf 1},2,1,{\bf 0},1,0,1)$. 
  
  Roughly speaking, the $(a,b)$-algorithm produces $Mod(H_{mix})$ by imposing\footnote{More details on "imposing" will be given in 6.1.2 for the similar $(a,c)$-algorithm. See also 7.5.} the $(a,b)$-wildcards one after the other. Notice that handling the last instance $(3,27,1)\ra (84,0)$ in Table 2 amounts to compressing the ideals of a poset. If we had chosen the $(a,b)$-algorithm then the 1’151’622 many 012-rows put forth by A-I-I would have given way to approximately 56'000 012ab-rows; see Table 4 in [W3] where a similar poset was evaluated.

  Why did we postulate that $H_{mix}$ be acyclic? That's because the (a,b)-algorithm has only been implemented for this scenario, and then enumerates [W2,Thm.5] the $N$ ideals of the underlying poset in time $O(Nw^2)$. In principle it extends to formulas $H_{mix}'$ in obvious ways (call it the {\it naive} (a,b)-algorithm) but becomes
   highly inefficient if the digraph $D(H'_{mix})$ (Sec.4.1) has large strong components. In the extreme case where it is strongly connected (which can be tested fast) we get $Mod(H'_{mix})=\{(0,0,...,0),(1,1,...,1)\}$, yet the naive  $(a,b)$-algorithm takes a long time to detect that.

 \vspace{5mm}
 {\bf 6.1.2} The somewhat\footnote{Boolean formulas of type $x_1\to(x_2\wedge x_3\wedge x_4)$ are sometimes (e.g. in the Data Mining community) called "implications". Among seven related types of "quasi-implication" features $x_1\to(\ol{x}_2\wedge \ol{x}_3\wedge \ol{x}_4)$. While psychologically all 8 types feel differently, logically they can be handled by just two algorithms. Either the (a,b)- or the (a,c)-algorithm, see [W2,p.133].} similar {\it (a,c)-wildcards}, like $(a,c,c,c):=(1,0,0,0)\uplus(0,2,2,2)$, are motivated by

 $$ (\ol{x}_1\vee\ol{x}_2)\vee(\ol{x}_1\vee\ol{x}_3)\vee(\ol{x}_1\vee\ol{x}_4)
\  \equiv\ \ol{x}_1\vee(\ol{x}_2\wedge \ol{x}_3\wedge \ol{x}_4)\
 \equiv\ x_1\to(\ol{x}_2\wedge \ol{x}_3\wedge \ol{x}_4)$$
 
\noindent
This leads [W4] to {\it 012ac-rows} and to some  {\it (a,c)-algorithm} that compresses the family $Acl(G)$ of all anticliques of $G$. In order to give an impression of its workings, consider the graph $G_2$ in Fig.5.

\begin{center}\includegraphics[scale=0.96]{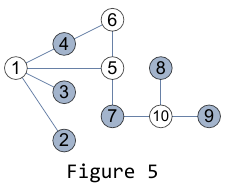}\end{center}

 The subset of vertices $VC=\{1,5,6,10\}$ is a {\it vertex-cover} of $G_2$, i.e. each edge of $G_2$ is incident with at least one vertex of $VC$. In order to get $Acl(G_2)$ it suffices to "impose" all wildcards $(a,c,..,c)$ where  $a$ ranges over $VC$ and the $c$'s match the neighborhood $NH(a)$.
 Hence in row $r_1$ of Table 5 the two vertices $1,10\in VC$ have already been "dealt with".
In order to handle $5\in VC$ which is pending for row  $r_1$, i.e. in order to sieve all $y\in r_1$ that satisfy $x_5\to(\ol{x}_1\wedge \ol{x}_6\wedge\ol{x}_7)$,  we define $r_{1,0}:=\{y\in r_1:\ y_5=0\}$ and $\ol{r}_{1,1}:=\{y\in r_1:\ y_5=1\}$. Obviously $r_1=r_{1,0}\uplus\ol{r}_{1,1} $.

\vspace{3mm}
\begin{tabular}{c|c|c|c|c|c|c|c|c|c|c|l}
	& 1 & 2 & 3& 4 & 5 & 6 & 7 & 8 &9 &10  & \\ \hline
	&       &       &    & &   &       &       &       &       &      & \\ \hline
	$r_1 =$ & $a_1$ & $c_1$ & $c_1$ &$c_1$ &$c_1$ &2 &$c_2$
         &$c_2$ &$c_2$ &$a_2$  & pending $5,6\in VC$   \\ \hline\hline

         $r_{1,0} =$ & $a_1$ & $c_1$ & $c_1$ &$c_1$ &{\bf 0} &2 &$c_2$
         &$c_2$ &$c_2$ &$a_2$  & pending $6\in VC$   \\ \hline

          $r_{1,1} =$ & 0 & 2 & 2 &2 &{\bf 1} &0 &0
         &$c_2$ &$c_2$ &$a_2$  &  final \\ \hline\hline

          $r_{1,0,0} =$ & $a_1$ & $c_1$ & $c_1$ &$c_1$ &0 &{\bf 0} &$c_2$
         &$c_2$ &$c_2$ &$a_2$  &  final \\ \hline

          $r_{1,0,1} =$ & $a_1$ & $c_1$ & $c_1$ &0 &0 &{\bf 1} &$c_2$
         &$c_2$ &$c_2$ &$a_2$  & final  \\ \hline
 \end{tabular}

 \vspace{2mm}
 {\sl Table 5: An impression of the standard (a,c)-algorithm}

\vspace{3mm}
\noindent
Evidently all members of $r_{1,0}$ satisfy $x_5\to(\ol{x}_1\wedge \ol{x}_6\wedge\ol{x}_7)$, and a moment's thought confirms that the set of all $y\in\ol{r}_{1,1}$ satisfying that formula is\footnote{In other words, $r_1\cap (c,2,2,2,a,c,c,2,2,2)=r_{1,0}\uplus r_{1,1}$, and so the wildcard $(a,c,c,c)$ (indexed by 5,1,6,7) has been {\it imposed} on $r_1$.} $r_{1,1}$ in Table 5. Incidently $r_{1,1}$ also "satisfies" $6\in VC$, and so $r_{1,1}$ is {\it final} in the sense that $r_{1,1}\s Acl(G_2)$. It remains to impose $6\in VC$ on $r_{1,0}$. One verifies that this yields $r_{1,0,0}$ and $r_{1,0,1}$.

 \vspace{5mm}
 {\bf 6.2} Seven previous (and less readible) versions of the present article exist in the arXiv. Despite  some worthy bits\footnote{Notably the "bisection factory" in Thm.2 of 
 arXiv:1208.2559v5 is an alternative to  A-I-I (and its extension in Sections 4,5) in that it also outputs $Mod(F)$ in polynomial total time, and compressed with 012-rows. But it is based on a digraph  more complicated than our $D(H')$ in Section 4. This complexity is inherited by the factor-poset which essentially is a "mirror-poset" in the sense of [CL]. Interestingly the bisection factory can be adapted to output all "partial models" w.r.t. a chosen window.  Related to partial models are {\it quantified} 2-CNFs (of which 'plain' 2-CNFs constitute the special case where all quantifiers are existential).}, the first five versions lack the crucial insight of Section 5 that each satisfiable 2-CNF can be switched to a Horn 2-CNF! 
 
 Why didn't we leave it at that and refer the reader to [W1] which shows how the modelset of any Horn CNF  can be enumerated? The answer: While Horn-CNFs are better than arbitrary CNFs, one should go on to exploit the special features of Horn 2-CNFs.

In order to flesh that out, observe that each satisfiable Horn CNF can be assumed\footnote{The  1-clauses can be "argued away" akin to the toy example in Section 1.} to have all its clauses  of type $\ol{x}_{i_1}\vee\cdots\vee\ol{x}_{i_t}\ (t\ge 2)$ or $\ol{x}_{i_1}\vee\cdots\vee\ol{x}_{i_t}\vee x_j\ (t\ge 1)$. Both shapes drastically simplify if all clauses are 2-clauses. The negative ones become $\ol{x}_i\vee\ol{x}_j$ and hence in combination model the anticliques of a graph. 
The mixed ones become $\ol{x}_i\vee x_j\ (\equiv x_i\to x_j)$ and hence in combination (upon factoring out the strong components) model the ideals of a poset. Graphs and posets are neat structures known to every mathematician. Many special types of graphs and posets are known for which all kinds of algorithms (including A-I-I?) accelerate.
There is another  difference between arbitrary Horn clauses and Horn 2-clauses.
In (R2) of Subsection 2.4 we had to test the feasibility of certain 012-rows $r''$ by checking whether some set $Y$ is an anticlique in some graph $G$. In contrast, in (14) of [W1] the evaluation of certain 012n-rows is clumsier and costs more time.

 \vspace{5mm}
 {\bf 6.3} Let us recap the story so far. You convinced yourself (hopefully) that in order to enumerate the models of an arbitrary 2-CNF $F$, it should first be "switched" to a Horn 2-CNF $H'$. Furthermore, 
 $H'$ should not undergo the (too general) treatment of Horn formulas in [W1]. Rather, upon transforming $H'$ to an {\it acyclic} Horn 2-CNF $H$, use A-I-I to optimally  enumerate $Mod(H)$  (from which $Mod(F)$ ensues immediately). 

 But then again, having read 6.1 and 6.2 the suspicion may arise that perhaps "optimally" is exaggerated. Viewing that $H'$ is of type $H'=H'_{mix}\wedge H_{neg}'$, why not reducing $H_{mix}'$ to $H_{mix}$ and $H_{neg}'$ to $H_{neg}$ as we did in Section 4? Afterwards calculate $Mod(H_{mix})$ and $Mod(H_{neg})$ with wildcards as shown in 6.1.
 Finally {\it combine the results} in order to get $Mod(H')$. Due to the additional wildcards the compression may be  higher than with A-I-I.
 
 In order to judge better (in 6.4) whether our suspicion is valid, we must clarify what is meant by "combine the results". First off, putting $H:=H_{mix}\wedge H_{neg}$ it holds (why?) that

 $$(8)\quad Mod(H)=Mod(H_{mix})\cap Mod(H_{neg}).$$

 In 6.3.1 we trim $Mod(H)$ to a more useful format, which then is used
 to get $Mod(H')$ in useful format. The latter happens in 6.3.2 and is easier.

  \vspace{5mm}
 {\bf 6.3.1}
  So suppose we found that $Mod(H_{mix})=\rho_1\uplus\cdots\uplus\rho_m$ with 012ab-rows $\rho_i$, and that $Mod(H_{neg})=\sigma_1\uplus\cdots\uplus\sigma_n$ with 012ac-rows $\sigma_j$. It then follows that

 $$(9)\quad Mod(H)=Mod(H_{mix})\cap Mod(H_{neg})=(\rho_1\cap\sigma_1)\uplus
 (\rho_1\cap\sigma_2)\uplus\cdots\uplus(\rho_m\cap\sigma_n).$$

\vspace{3mm}
\noindent
Each one of the $mn$ intersection $\rho_i\cap\sigma_j$ may or may not expand smoothly to a more useful format. Here come\footnote{In order to avoid ambiguity between $(a,b,..,b)$ and $(a,c,..,c)$ when they appear simultaneously, we henceforth replace the latter by $(\underline{a},c,..,c)$.} some concrete instances.

\vspace{3mm}
\begin{tabular}{c|c|c|c|c|c|c|c|c|c|c|c|l}
	& 1 & 2 & 3& 4 & 5 & 6 & 7 & 8 &9 &10 & 11\\ \hline
	&       &       &    & &   &  &     &       &     &     &      & \\ \hline
	$\rho_2 =$ & $a_1$ &$b_1$ & $b_1$ & $a_2$ &$b_2$ &$b_2$ &$a_3$ &$b_3$
         &$b_3$ &$a_4$ &$b_4$    \\ \hline
         
         $\sigma_5 =$ & $2$ &$2$ & $0$ & $2$ &$1$ &$2$ &$0$ &$2$
         &$2$ &$1$ &$2$    \\ \hline

         $\rho_2\cap\sigma_5 =$ & $0$ &$2$ & $0$ & $a_2$ &$1$ &$b_2$ &$0$ &$2$
         &$2$ &$1$ &$1$    \\ \hline\hline

          $\rho_7 =$ & $2$ &$2$ & $0$ & $2$ &$1$ &$2$ &$0$ &$2$
         &$2$ &$1$ &$2$    \\ \hline

         $\sigma_4 =$ & $a_1$ &$c_1$ & $c_1$ & $a_2$ &$c_2$ &$c_2$ &$a_3$ &$c_3$
         &$c_3$ &$a_4$ &$c_4$    \\ \hline

           $\rho_7\cap\sigma_4 =$ & $a_1$ &$c_1$ & $0$ & $0$ &$1$ &$2$ &$0$ &$2$
         &$2$ &$1$ &$0$    \\ \hline\hline

         $\rho_8 =$ & $a_1$ &$b_1$ & $b_1$ & $2$ &$0$ &$2$ &$b_2$ &$b_2$
         &$b_2$ &$b_2$ &$a_2$    \\ \hline

         $\sigma_3 =$ & $2$ &$2$ & $1$ & $\underline{a}_1$ &$c_1$ &$c_1$ &$c_1$ &$c_2$
         &$\underline{a}_2$ &$0$ &$c_2$    \\ \hline

         $\rho_8\cap\sigma_3 =$ & $a_1$ &$b_1$ & $1$ & $\underline{a}_1$ &$0$ &$c_1$ &$c_1$ &$c_2$ &$\underline{a}_2$ &$0$ &$0$    \\ \hline\hline

         $\rho_5 =$ & $a_1$ &$b_1$ & $b_1$ & $a_2$ &$b_2$ &$b_2$ &$b_2$ &$a_3$
         &$b_3$ &$b_3$ &$b_3$    \\ \hline

          $\sigma_6 =$ & $c_1$ &$\underline{a}_3$ & $c_2$ & $\underline{a}_2$ &$\underline{a}_4$ &$\underline{a}_1$ &$c_4$ &$c_4$
         &$c_1$ &$c_2$ &$c_3$    \\ \hline
 \end{tabular}

 \vspace{3mm}
 {\sl Table 6: About intersecting 012ab-rows with 012$\underline{a}c$-rows}

 \vspace{3mm}
 Generalizing $\rho_2\cap\sigma_5$, intersections of 012-rows with 012ab-rows are always either empty or can be recast as 012ab-rows. Generalizing $\rho_7\cap\sigma_4$, intersections of 012-rows with 012ac-rows are always either empty or can be recast as 012ac-rows. As to the intersection of a proper 012ab-row $\rho$ with a proper 012$\underline{a}$c-row $\sigma$, either one, say $\sigma$, can by 6.1.1 be expanded as disjoint union of $2^t$ 012-rows. It follows that $\rho\cap\sigma$ can be expanded as disjoint union of at most $2^t$ many 012ab-rows. We call this the {\it naive way}. If one is lucky one can do better
 (see $\rho_8\cap\sigma_3$) and only use one or few {\it $012ab\underline{a}c$-rows} (i.e. rows that may feature both (a,b)- and (a,c)-wildcards). 
 As to "few", no research\footnote{Readers are encouraged to help out. Here comes a rough line of attack to expand $\rho\cap\sigma$ as disjoint union of few $012ab\underline{a}c$-rows. Namely, suppose $\rho(0,1)$ originates from $\rho$ by setting two components to 0 and 1. They must be well-chosen in  that $\rho(0,1)\cap\sigma$ expands as a {\it single} $012ab\underline{a}c$-row. So far, so good. But it remains to rewrite $\rho(0,0)\cap\sigma$ and $\rho(1,0)\cap\sigma$ and $\rho(1,1)\cap\sigma$. Hence repeat the procedure by finding  well-chosen positions in $\rho(i,j)$ or $\sigma$. And so forth. } on this intruiging matter has been undertaken.

\vspace{5mm}
 Therefore in the remainder we rather strive for an {\it unbalanced scenario} (the more extreme the better) in the sense of having many 012ab-rows and few 012ac-rows, or the other way around. As to "strive", in view of 5.2.1 and 5.2.2 unbalancedness in favor of $H_{mix}'$ or $H_{neg}'$ can be influenced to some extent. This unbalancedness probably will be reflected in $H=H_{mix}\wedge H_{neg}$. 

 To illustrate the benefit of unbalancedness, consider two scenarios taking place in (9).\\
 {\it Case 1 (balanced):} $m=n=100$. Then $mn=10000$ intersections $\rho_i\cap\sigma_j$ must be dealt\footnote{Be it in some clever way, be it in the naive way.} with. {\it Case 2 (unbalanced):} $m=180,\ n=20$. Then only $mn=3600$ intersections  must be dealt with.

 \vspace{3mm}
 {\bf 6.3.2} We thus managed to write $Mod(H)$ as a disjoint union of $h$ many
 $012ab\underline{a}c$-rows $\tau_k$ which in turn originated upon processing the $mn$ intersections $\rho_i\cap\sigma_j$ in (9). The intuition is that $h$ is larger than $mn$, but not necessarily so if many intersections are empty. If we were forced (using, say, Mathematica) to code  "process all $\rho_i\cap\sigma_j$", we would have to employ, for the time being, the naive way.
 Thus imagine that our toy row $\tau_0$ in Table 7 also originated in this way, i.e. $\tau_0=\rho_0\cap\sigma_0$, where $\rho_0$ is some 012ab-row  and $\sigma_0$  some 012-row. 

 \vspace{2mm}
 \begin{tabular}{c|c|c|c|c|c|c|c|c|c|c|l}
	& 1 & 2 & 3& 4 & 5 & 6 & 7 & 8 &9 &10  & \\ \hline
	&       &       &    & &   &       &       &   
    &       &      & \\ \hline
	$\rho_0 =$ & $b_1$ & $a_2$ & $b_1$ &$a_1$ &$b_2$ &$b_1$ & &
    & &     &   \\ \hline

    $\sigma_0 =$ & $2$ & $0$ & $1$ &$2$ &$2$ &$2$ & &
    & &     &   \\ \hline

$\tau_0 =$ & $b_1$ & $0$ & $1$ &$a_1$ &$2$ &$b_1$ & &
    & &     &   \\ \hline

$\ol{\tau_0} =$ & $b_1$ & $0$ & $1$ &$a_1(i)$ &$2(i)$ &$b_1(i)$ &$2(i)$ &$1$
    &$a_1(i)$ &$b_1(i)$    &   \\ \hline
    
\end{tabular}

\vspace{3mm}
{\sl Table 7: Extending $012ab\underline{a}c$-rows $\tau_0$ (degenerate or not)  to
$012ab\underline{a}c(i,ii)$-rows $\ol{\tau_0}$ }

\vspace{3mm}
Recall that the bitstrings in $Mod(H)$ are in bijection with the "good" ideals $Y$ of a poset $P$, where "good" means that $Y$ is simultaneously an anticlique of some graph sharing the same universe $\{1,2..,6\}$ as $P$. Unfortunately $P$ is the poset of strong components some digraph, say with vertex-set $\{1,2,...,10\}$ (never mind the bad filter). Say the strong components are $\{1\},\{2\},\{3,8\},\{4,9\},\{5,7\},\{6,10\}$. Then $\tau_0$ triggers the row $\ol{\tau_0}$ (Table 7) which is a subset of $Mod(H')$. We call $\ol{\tau_0}$ and 
the likes\footnote{As another example, an indexed (a,b)-wildcard, say $(a_7,b_7,b_7,b_7,b_7)$, whose entries belong to strong components of cardinalities 3,2,1,3,1, becomes 
$(a_7({\rm i}),a_7({\rm i}),a_7({\rm i}),\ b_7({\rm i}),b_7({\rm i}),\ b_7,\ b_7({\rm ii}),b_7({\rm ii}),b_7({\rm ii}),\ b_7)  $. We use i,ii,iii,.. instead of 1,2,3,.. because Arabic numbers and indices abound already. } 012ab(i,ii)-rows. They constitute a special case of
$012ab\underline{a}c(i,ii)$-rows.

 \vspace{5mm}
 {\bf 6.4} We are now better equipped to judge the suspicion that came up in 6.3. First, recall the standard recipee layed out in Sections 2,4,5: Given a satisfiable 2-CNF $F$, get a matching Horn 2-CNF $H'$, then turn it into an acyclic $H$, then get $Mod(H)$ with A-I-I, then get $Mod(F)$ from $Mod(H)$. The  suspicion was
 that "get $Mod(H)$ with A-I-I"  may be challenged by "get $Mod(H)$ by combining the (a,b)- with the (a,c)-algorithm".
As seen, this suspicion is indeed valid, the more so the more unbalanced our scenario is. 

 \vspace{2mm}
 \begin{tabular}{c|c|c|c|c|c|c|c|c|c|c|l}
	& 1 & 2 & 3& 4 & 5 & 6 & 7 & 8 &9 &10  & \\ \hline
	&       &       &    & &   &       &       &       &       &      & \\ \hline
    $\ol{\tau_{0}}$ & $b_1$ & $0$ & $1$ &$a_1(\rm{i})$ &$2(\rm{i})$ &$b_1(\rm{ii})$ &$2(\rm{i})$ &$1$ &$a_1(\rm{i})$ &$b_1(\rm{ii})$  &   \\ \hline  

    $\Theta=!!\ol{\tau_{0}}:=$ & $b_1$ & $0\ !!$ & $1$ &$a_1(\rm{i})$ &${2(\rm{i})}\ !!$ &$b_1(\rm{ii})$ &$2(\rm{i})$ &$1$ &${a_1(\rm{i})}\ !!$ &$b_1(\rm{ii})$  &   \\ \hline 
    &       &       &    & &   &       &       &       &       &      & \\ \hline

$\ol{\tau_{0,1}}:=$ & $2$ & $0$ & $1$ &$\bf 0$ &$0$ &$2(\rm{ii})$ &$0$ &$1$ &$\bf 0$ &$2(\rm{ii})$  &   \\ \hline

$\ol{\tau_{0,2}}:=$ & $2$ & $0$ & $1$ &$\bf 0$ &$1$ &$2(\rm{ii})$ &$1$ &$1$ &$\bf 0$ &$2(\rm{ii})$  &   \\ \hline

 $\ol{\tau_{0,3}}:=$ & $1$ & $0$ & $1$ &$\bf 1$ &$0$ &$1$ &$0$ &$1$ &$\bf 1$ &$1$  &   \\ \hline

  $\ol{\tau_{0,4}}:=$ & $1$ & $0$ & $1$ &$\bf 1$ &$1$ &$1$ &$1$ &$1$ &$\bf 1$ &$1$  &   \\ \hline

 &       &       &    & &   &       &       &       &       &      & \\ \hline

  $\Theta_1:=$ & $2$ & $\bf 1$ & $1$ &$0$ &$\bf 1$ &$2(\rm{ii})$ &$0$ &$1$ &$\bf 1$ &$2(\rm{ii})$  &   \\ \hline

$\Theta_2:=$ & $2$ & $\bf 1$ & $1$ &$ 0$ &$\bf 0$ &$2(\rm{ii})$ &$1$ &$1$ &$ \bf 1$ &$2(\rm{ii})$  &   \\ \hline

 $\Theta_3:=$ & $1$ & $\bf 1$ & $1$ &$ 1$ &$\bf 1$ &$1$ &$0$ &$1$ &$\bf 0$ &$1$  &   \\ \hline

  $\Theta_4:=$ & $1$ & $\bf 1$ & $1$ &$ 1$ &$\bf 0$ &$1$ &$1$ &$1$ &$\bf 0$ &$1$  &   \\ \hline   
\end{tabular}

\vspace{3mm}
{\sl Table 8: Unpacking the 012!!ab(i,ii)-row $\Theta$}
\vspace{3mm}

Unfortunately there still remains the task to "get $Mod(F)$ from $Mod(H)$".
Actually, by 6.3.2, we already reduced the task to "get $Mod(F)$ from $Mod(H')$".
Recall, $Mod(H')$ is now a disjoint union of $012ab\underline{a}c(i,ii)$-rows. To fix ideas, let $\ol{\tau_0}$ from Table 7 be one of these rows.
Suppose the variables that were switched (long time ago in order to get $H'$ from $F$) were $x_2,x_5,x_9$. 
What, then, are the bitstrings in $Mod(F)$ that match the bitstrings in $\ol{\tau_0} \s Mod(H')$? Let $S\s Mod(F)$ be that set of  bitstrings. A bold representation of $S$ is $!!\ol{\tau_0}$ where !! indicates which variables have been switched. This is good and well, but how to unpack $!!\ol{\tau_0}$ if a more explicite\footnote{Unpacking is not necessary for merely getting cardinalities: $|!!\ol{\tau_0}|=|\tau_0|=|(b_1,0,1,a_1,2,b_1,1)|=10$.} format is required?

One checks that $\ol{\tau_{0}}=\ol{\tau_{0,1}}\uplus\cdots\uplus\ol{\tau_{0,4}}$, the decomposition being based on whether $(a_1(\rm{i}),a_1(\rm{i}))$ equals $(0,0)$ or
$(1,1)$, and on whether $(2(\rm{i}),2(\rm{i}))$ equals $(0,0)$ or
$(1,1)$. In general the decomposition must be such that the entries with the indices of the switched variables must all be 0,1, or 2.  By definition $\Theta_i$ arises from
$\ol{\tau_{0,i}}$ upon switching the variables $x_2,x_5,x_9$.
It follows that $\Theta=\Theta_1\uplus\Theta_2\uplus\Theta_3\uplus\Theta_4$.

\vspace{4mm}
 {\bf 6.4.1} So, at the end of the day, is it worthwile to embark on wildcards fancier than the don't-care symbol? The following five scenarios are the most clear-cut. 
 
 First, one should not embark on wildcards if $Mod(F)$ is small (which can often be predicted by e.g. using Mathematica's {\tt SatisfyabilityCount}) since then launching the machinery of wildcards hardly pays off with extra compression. 
 
 Second, it is mostly  a good idea to use (a,c)-wildcards when
 the 2-CNF $F$ is {\it homogeneous} in the sense that all its clauses are exclusively  negative.
 There are no strong components to worry about and according to Thm. 1 in [W4] the (a,c)-algorithm returns the $N$ anticliques in time $O(Nw^2)$, where $w$ is the number of variables in $F$.

 Third, what if $F$ has exclusively positive clauses? As one may speculate, this works "dually" to the above. Spelling it out, let $F'$ be the 2-CNF obtained from $F$ by switching {\it all} variables. Then $F'$  has exclusively negative clauses, and so $Mod(F')$ expands as a disjoint union of 012ac-rows. Upon replacing them by 012-rows and then switching 0's and 1's, one obtains $Mod(F)$. One can in fact retain the compression achieved in $Mod(F')$
 by replacing each wildcard $(a,c,c,c)$ (recall that's $(1,0,0,0)\uplus(0,2,2,2)$) by the novel wildcard $(\alpha,\gamma,\gamma,\gamma):=(0,1,1,1)\uplus (1,2,2,2)$. One also could apply right away some obvious $(\alpha,\gamma)$-algorithm to some obvious graph\footnote{The edges of this graph $G$ match the clauses $x_i\vee x_j$ of $F$, and $Mod(F)$ matches the set of all vertex covers of $G$.} defined by $F$.

Fourth, it is mostly a good idea to use (a,b)-wildcards when
 the 2-CNF $F$ has merely mixed clauses, {\it provided} $F=H_{mix}$ is acyclic. As mentioned before, then the (a,b)-algorithm\footnote{As to enhancing the (a,b)-algorithm, readers are encouraged to carry out the idea of [W2,p.134].} returns  $Mod(H_{mix})$ as disjoint union of 012ab-rows. 
But what if $F=H'_{mix}$ is cyclic? This could be handled by the naive (a,b)-algorithm, which has not been implemented yet. Recall that it would be the more inefficient the larger the strong components are. On the plus side, lifting 012ab-rows to 012ab(i,ii)-rows does not take place. In fact, in some cases (such as below) aiming for 012-rows instead of 012ab-rows is preferable.

\vspace{3mm}
Last not least, here comes a variant which doesn't come with a time bound, but it is "clean" and represents $Mod(F)$ as disjoint union of good old 012ac-rows for {\it any} satisfiable 2-CNF $F$ (with set of variables $W$). Namely, upon Horn-renaming switch to $H'=H'_{mix}\wedge H'_{neg}$. Using the {\it naive} (a,b)-algorithm, we can stick with $W$ (thus no factor-posets) and represent $Mod(H'_{mix})$ as disjoint union of 012-rows $\rho_i$. Since $Mod(H'_{neg})$ is a disjoint union of 012ac-row (upon applying the standard (a,c)-algorithm), all nonempty intersections $\rho_i\cap\sigma_j$ are (fast calculated) 012ac-rows, and their disjoint union equals $Mod(H')$. The latter (Table 8) immediately yields $Mod(F)$.

\vspace{3mm}
{\bf 6.5} A general remark on wildcards and output-polynomial enumeration. For general Horn CNFs output-polynomial enumeration of the modelset can be achieved with or without wildcards [W1]. For {\it general} 2-CNFs output-polynomiality can be achieved (be it with [F] or with A-I-I + Sections 4,5) only {\it without} wildcards. For all other CNFs  output-polynomiality cannot even be achieved with 01-rows (i.e. one-by-one enumeration). 

\section{The  All-SAT landscape in a nutshell}

The reference  arXiv:1208.2559v5 given in 6.2 discusses an idea (prior to A-I-I) for obtaining the modelset of a 2-CNF $F$. Furthermore  arXiv:1208.2559v5 discusses Feder's algorithm [F] from the introduction. It seems the latter has never been implemented.
There are other methods but they apply to {\it arbitrary} Boolean functions $F$ and thus fit the general {\it All-SAT} problem\footnote{ As opposed to SAT, which finds (if existing) {\it one} model of $F$, All-SAT produces {\it all} of them. One reviewer
admonished not only a survey of the All-SAT landscape (which occurs in Sec.7) but also a comparison of A-I-I with "state of the art" All-SAT algorithms different from {\tt BooleanConvert}. This is a noble cause for the future but I can participate in it at most as co-author, having neither time, nor capability doing it on my own. }. In the remainder we touch upon five  All-SAT methodologies. 

\vspace{5mm}
{\bf 7.1} The most natural approach is {\it Pivotal Decomposition}. Here one chooses a suitable, or any,  variable (=pivot) $x_i$ and sets it to $0$ or $1$. This gives

$$Mod(F)=\{x\in Mod(F):\ x_i=0\}\uplus\{x\in Mod(F):\ x_i=1\}=:M_0\uplus M_1.$$

\noindent
One proceeds likewise to $M_{00},M_{01},M_{10},M_{11}$, and generally to $M_{ij...k}$. This requires a SAT-solver to immediately discard empty sets $M_{ij...k}$.
Furthermore one stops splitting $M_{ij...k}$ as soon as it can be written as 012-row. Pivotal Decomposition works for {\it arbitrary} Boolean formulas, such as $F:=[(x_2\vee x_3)\to x_1]\wedge [(x_3\vee x_4)\to (x_1\wedge x_2)]$ which has [W3,p.1075] modelset  $Mod(F)=(1,1,2,2)\uplus(1,0,1,0)\uplus(2,0,0,2)$.

Pivotal Decomposition has the advantage that the pivots
can be chosen on the fly. To spell this out, most sets $M_{ij...k}$ are unknown but each is accompanied by a  012-row $r_{ij...k}\supseteq M_{ij...k}$. Furthermore one carries along the Boolean formula
 $F_{ij...k}$, whose models are the bitstrings in $M_{ij...k}$, and which hence needs to be "imposed" on $r_{ij...k}$.
 The formulas $F_{ij...k}$ are obtained inductively from the original formula $F$ by substituting variables with 0's and 1's. Suppose $r_{ij...k}$ is on top of the LIFO stack.
 Using an appropriate heuristic one  decides "on the fly" which pivot would split $F_{ij...k}$ into the two most convenient new formulas. (There is no way predicting the best order of pivots at the outset!). Whenever $F_{ij...k}$ is a tautology, its matching 012-row $r_{ij...k}\ (=M_{ij...k}$ here) is removed from the stack and becomes part of $Mod(F)$.

Pivotal Decomposition works particularly well\footnote{In fact one needs not wait until $F_{ij...k}$ is a tautology; using the right pivots, $M_{ij...k}$ often becomes an obvious   {\it union} of 012-rows. Interestingly, CNF's behave nicely in this regard as well (work in progress).} when $F$ is in DNF-format. In fact it all began with DNF's  in the 80's in the context of computing  reliability polynomials of  networks. See [W3] and its references.

\vspace{5mm}
{\bf 7.2} In 1986 Randal Bryant launched the first glorious decade of {\it Binary Decision Diagrams (BDDs)}. Here is what a BDD looks like:

\begin{center}\includegraphics[scale=0.8]{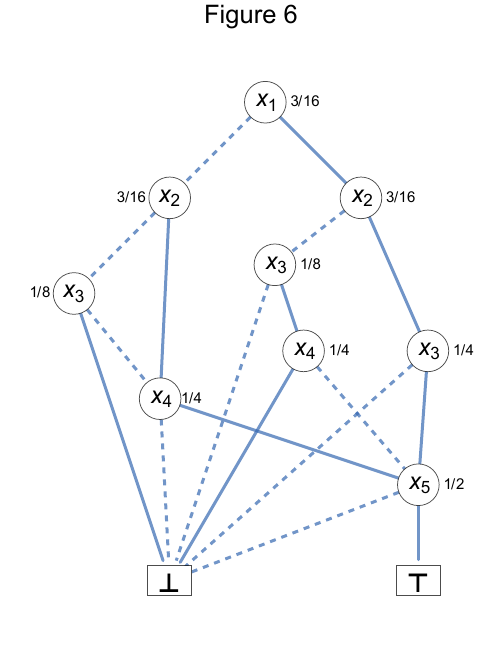}\end{center}

The dashed and solid lines departing from a node $x_i$ are chosen according to whether $x_i=0$ or $x_i=1$. Hence each bitstring $x=(x_1,...,x_5)$ determines a path from the {\it root} (=top node) $x_1$ to $\tt T$ (=true) or $\perp$ (=false). For instance $(0,1,0,1,1)$ yields $\tt T$, while each bitstring in $(1,0,0,2,2)$ yields $\perp$. Hence the BDD in Fig.6 insinuates a Boolean function $F_{BDD}$ which by definition has $x\in Mod(F_{BDD})$ iff $x$ yields $\tt T$.
Conversely, given a Boolean formula $F$ {\it and} a fixed ordering of the variables (say $x_4,x_1,x_5,x_3,x_2$) there is a unique binary decision diagram $BDD(F)$ (with root $x_4$) that  {\it represents} $F$ (thus $Mod(F)=Mod(F_{BDD(F)})$. While it is hard calculating $BDD(F)$ (see [K]), many rewards can be reaped upon success; let us glimpse at three. 

First, one gets a compression of $Mod(F)$ by  enumerating all paths from the root to $\tt T$ (which is easy). For instance for $BDD(F)$  in Fig.6 we saw already that $(0,1,0,1,1)\in (0,1,2,1,1)\s Mod(F)$. By inspection one finds that altogether there are four "good" paths and they yield $Mod(F)=(0,1,2,1,1)\uplus(0,0,0,1,1)\uplus(1,0,1,0,1)\uplus(1,1,1,2,1)$. In particular $|Mod(F)|=2+1+1+2=6$. Second, let $F_1,F_2$ be two (possibly quite different) formulas. Then $F_1$ and $F_2$ are {\it equivalent} (i.e. have the same modelset) iff $BDD(F_1)=BDD(F_2)$. This can be decided  faster than computing the modelsets themselves. Third, given $BDD(F)$ the cardinality $|Mod(F)|$ can also be found {\it without calculating} $Mod(F)$ as follows\footnote{The Mathematica command {\tt SatisfiabilityCount} used in Section 3 exploits this idea.}. Upon labeling the nodes of the BDD from bottom to top with rational numbers in an obvious recursive way (which is a variant of Algorithm C in [K,p.207]) the top number is the probability that a random bitstring  evaluates to $\tt T$. In Fig.6 this probability is $\frac{3}{16}$, and so $|Mod(F)|=2^5\cdot \frac{3}{16}=6 $
(as previously seen).

One downside of BDD's is that even when $BDD(F)$ is reasonably small, the intermediate BDDs occuring during calculation may be excessively large. This is also due to the fact that determining a good (let alone optimum) variable order is hard, and cannot be done on the fly as in 7.1. This (and the competition in 7.3) led to what Minato calls the BDD-winter from 1999 to 2005. But BDD's have resurfaced from hibernation, because they found their niche and also morphed to so-called zero-supressed decision diagrams (ZDD's); see [M].

\vspace{5mm}
{\bf 7.3} The BDD-winter was in part caused by {\it SAT+BC}. This All-SAT strategy combines SAT with so-called "blocking clauses". We illustrate the method on the Boolean function $H_1$ in (4). The first step is to use your SAT-solver of choice to find any model of $H_1$, say $r_1':=(1,1,0,1,0,0,0,1)$. Next we inflate $r'_1$ by replacing as many as possible bits with don't-cares while staying within $Mod(H_1)$. This e.g. yields
$r'_2:=(1,2,0,2,0,0,0,2)\s Mod(H_1)$. In order to prevent a future repetition of models we introduce the {\it blocking clause} $C_2:=\ol{x}_1\vee x_3\vee x_5\vee x_6\vee x_7$. It is evident that $Mod(H_1\wedge C_2)=Mod(H_1)\setminus r_2'$. Next our SAT-solver finds that
say $r_3:=(1,0,1,0,1,0,0,0)$ satisfies $H_1\wedge C_2$. By inflating $r_3'$ we e.g. obtain
$r_4':=(1,0,1,0,2,0,0,2)\s Mod(H_1)$. The blocking clause $C_4$ that matches $r_4'$ achieves
that $Mod(H_1\wedge C_2\wedge C_4)=Mod(H_1)\setminus (r_2'\uplus r_4')$.

\vspace{3mm}
\begin{tabular}{c|c|c}
    Applying SAT & inflating $r_i'$ to $r_{i+1}'$ & blocking clause for $r_{i+1}'$ \\ \hline\hline
    $r_1'=(1,1,0,1,0,0,0,1)$ &  $r_2'=(1,2,0,2,0,0,0,2)$ & $C_2:=\ol{x}_1\vee x_3\vee x_5\vee x_6\vee x_7$ \\ \hline  

     $r_3'=(1,0,1,0,1,0,0,0)$ &  $r_4'=(1,0,1,0,2,0,0,2)$ & $C_4:=\ol{x}_1\vee x_2\vee\ol{x}_3\vee x_4\vee x_6\vee x_7$ \\ \hline 

     $r_5'=(0,1,0,1,0,1,0,0)$ &  $r_6'=(0,1,0,2,0,2,0,0)$ & $C_6:=x_1\vee \ol{x}_2\vee x_3\vee x_5\vee x_7\vee x_8$ \\ \hline 

     $r_7'=(1,1,0,1,0,1,0,0)$ &  $r_8'=(1,1,0,2,0,1,0,0)$ & $C_8:=\ol{x}_1\vee\ol{x}_2\vee x_3\vee x_5\vee\ol{x}_6\vee x_7\vee x_8$ \\ \hline 

     $r_9'=(0,0,0,1,0,0,0,0)$ &  $r_{10}'=(0,0,0,2,0,0,0,0)$ & 
     $C_{10}:=x_1\vee x_2\vee x_3\vee x_5\vee x_6\vee x_7\vee x_8$ \\ \hline

      $r_{11}'=(0,0,1,0,1,0,0,0)$ &  $r_{12}'=(0,0,1,0,2,0,0,0)$ & 
      
     $C_{12}:=x_1\vee x_2\vee \ol{x}_3\vee x_4\vee x_6\vee x_7\vee x_8$ \\ \hline
     
  \end{tabular}

\vspace{3mm}
{\sl Table 9: The All-SAT algorithm SAT+BC in action}

\vspace{3mm}
And so it goes on (see Table 9) until one finds that $H_1\wedge C_2\wedge\ldots\wedge C_{12}$ is insatisfiable. This amounts to $Mod(H_1)=r_2'\uplus r_4'\uplus\cdots\uplus r_{12}'$.
With a little effort\footnote{Checking $|r_4'|+\cdots+|r_{12}'|=24=|r_6|+\cdots+|r_{11}|$ is easier.} one verifies that indeed $r_2'\uplus r_4'\uplus\cdots\uplus r_{12}'=r_6\uplus r_7\uplus r_8\uplus r_{10}\uplus r_{11}$, where the latter rows are from Table 1.
While SAT+BC caters for arbitrary Boolean formulas, it excels when $F$ is in CNF-format because then inflating 01-rows (=models found by SAT) to fat 012-rows amounts to cleverly solve  a certain set covering problem (and this task is well researched). One disadvantage of SAT+BC is that the more blocking clauses are added, the more the SAT-solver struggles to find models of the increasingly longer formulas.

\vspace{5mm}
{\bf 7.4} In line with [TS] we label the fourth All-SAT strategy  {\it SAT+NBC}. While NBC stands for "no blocking clauses" and hence emphasizes the difference to 7.3, here comes   {\it what replaces} the blocking clauses. Other than in 7.1 and 7.3 the "SAT" in "SAT+NBC"
refers to a {\it particular} SAT-solver, i.e. the well-known DPLL method (more in 7.6). Essentially SAT+NBC is the iteration of "DPLL with a twist".
This twist concerns the chronological backtracking ingredient of DPLL and ensures that upon iteration of DPLL no models are duplicated. We note that in the simplest version of SAT+NBC
the models are output [TS, top of page 1.12:13] one-by-one (in contrast to SAT+BC).

\vspace{5mm}
{\bf 7.5} While the All-SAT methods presented so far ignore the format of $F$ (albeit 7.1 has a propensity for DNFs, and 7.4 for CNFs), the fifth method  {\it demands} CNF-format. Given $F=C_1\wedge C_2\wedge\ldots \wedge C_h$ the clauses $C_i$ are "imposed" one after the other until $Mod(F)$ is reached:

$$Mod(C_1)\supseteq Mod(C_1\wedge C_2)\supseteq\cdots\supseteq Mod(C_1\wedge\ldots\wedge C_{h-1})\supseteq Mod(F)$$

\noindent
The  (a,b)-algorithm and (a,c)-algorithm are instances of this general idea of {\it Clausal Imposition}. In fact, recall from Subsection 6.1 that both algorithms have the perk that certain groups of 2-clauses $C_i$ (i.e. sharing a common literal) can be imposed all at once.
Many more instances of Clausal Imposition are surveyed in [W3]. All of them have their own taylor-made wildcards beyond the don't-care symbol.

\vspace{5mm}
{\bf 7.6}  A word on {\tt BooleanConvert} from Section 3. It is a blackbox All-SAT algorithm; the author failed to find out about its inner workings. Most readers will agree that the {\it crispest} of the discussed  All-SAT strategies are 7.1 and 7.3. As stated in [TS], the {\it most popular} ones are indeed 7.2, 7.3, 7.4. Except for BDDs, all discussed strategies require a SAT-solver as a subroutine.
This leads us to the Handbook of Satisfiability [BHMW]. The acronym DPLL used in 7.4 stands for Davis-Putnam-Logemann-Loveland and their algorithm frequently occurs  in [BMHW]. And so do lesser known SAT-solvers, a luscious historical account (starting with Aristotle), random satisfiability, and much more. Various Applications+Extensions of SAT  feature in [BMHW, Part 2], such as MaxSAT, Cardinality constraints, Quantified Boolean functions. All of this is great. On the downside, All-SAT does {\it not}\footnote{Specifically, the word all-SAT is only  mentioned once on page 808. While BDD's pop up a few times, their All-SAT capability (see 7.2) is only briefly touched upon on page 31 and 107.} feature among these applications.   The application that comes closest to All-SAT is chapter 25 of [BHMW], i.e. Model Counting. As glimpsed in 7.2, counting models is easier than generating them, and uses other lines of attack.

\section{References}

\begin{enumerate}
	
		 \item[\bf BHMW]   A.Biere, M. Heule, H. van Maaren, T. Walsh (editors), The Handbook of Satisfiability, IOS Press BV 2021.

         \item[\bf CH] Y. Crama, P.L. Hammer, Boolean functions, Encyc. Math. and Appl. 142, Cambride University Press 2011.

     \item[\bf CL] CT Cheng, A Lin, Stable roommates matchings, mirror posets, median graphs, and the local/global median phenomenon in stable matchings, SIAM J. Discrete Mathematics, 25 (2011) 72-94.

    \item[\bf CLM] N. Caspard, B. Leclerc, B. Monjardet, Finite Ordered Sets, Cambridge University Press 2012.
	
	\item[\bf F] T. Feder, Network flow and 2-satisfiability, Algorithmica 11 (1994) 291-319.
	\item[\bf FK] M. F\"{u}rer, S.P. Kasiviswanathan, Algorithms for counting 2-SAT solutions and colorings with application, Lecture Notes in Computer Science 4508 (2007) 47-57.

     \item[\bf K] D.E. Knuth, The Art of Computer Programming, Volume 4A, Addison-Wesley 2012.

      \item[\bf LXCCW] Y. Lin, Z. Xie, T. Chen, X. Cheng, H. Wen, Image privacy protection scheme based on high-quality reconstruction DCT
compression and nonlinear dynamics, Expert Systems With Applications 257 (2024) 124891.

      \item[\bf Ma] D. Manlove, Algorithmics Of Matching Under Preferences, World Scientific 2013.

      \item[\bf M] S. Minato, Power of Enumeration — Recent Topics on BDD/ZDD-Based
Techniques for Discrete Structure Manipulation, IEICE transactions on information and systems 100 (8), 1556-1562.

         \item[\bf RW] C. Rohwer, M. Wild, LULU Theory, idempotent stack filters, and the mathematics of vision of Marr, Advances in Imaging and Electron Physics 146 (2007) 57-162.
         
	\item[\bf S] B. Simeone, Quadratic functions, Chapter 5 in [CH].

     \item[\bf TS] T.Toda, T. Soh, Implementing Efficient All Solutions SAT Solvers,
     ACM Journal of Experimental Algorithmics, Vol. 21, No. 1, Article 1.12, 2016.

	\item[\bf W1] M. Wild, Compactly generating all satisfying truth assignments of a Horn formula, Journal on Satisfiability, Boolean Modeling and Computation 8 (2012) 63-82.
	\item[\bf W2] M. Wild, Output-polynomial enumeration of all fixed-cardinality ideals of a poset, respectively all fixed-cardinality subtrees of a tree, Order 31 (2014)  121-135.
    \item[\bf W3] M. Wild, Compression with wildcards: From CNFs to orthogonal DNFs by imposing the clauses one-by-one, The Computer Journal 65 (2022) 1073-1087.
		\item[\bf W4] M. Wild, ALLSAT compressed with wildcards: All anticliques (or cliques), arXiv:0901.4417v6

\end{enumerate}

\end{document}